\documentclass[a4paper,fleqn,usenatbib]{mnras}

\usepackage{newtxtext,newtxmath}

\usepackage[T1]{fontenc}
\usepackage{ae,aecompl}

\usepackage{graphicx}	
\usepackage{amsmath}	
\usepackage{amssymb}	

\title[Magnetic field orientation in molecular clouds]{From parallel to perpendicular -- On the orientation of magnetic fields in molecular clouds}
  \author[D. Seifried et al. ]
  {D.~Seifried,$^1$\thanks{seifried@ph1.uni-koeln.de} S. Walch,$^1$ M. Weis,$^1$ S. Reissl,$^{2,3}$ J. D. Soler,$^4$ R. S. Klessen,$^{2,3}$ P. R. Joshi$^1$ \\
  $^1$Universit\"at zu K\"oln, I. Physikalisches Institut, Z\"ulpicher Str. 77, 50937 K\"oln, Germany\\
  $^2$Universit\"{a}t Heidelberg, Zentrum f\"{u}r Astronomie, Institut f\"{u}r Theoretische Astrophysik, Albert-Ueberle-Str. 2, 69120 Heidelberg, Germany\\
  $^3$Universit\"{a}t Heidelberg, Interdisziplin\"{a}res Zentrum f\"{u}r Wissenschaftliches Rechnen, Im Neuenheimer Feld 205, 69120 Heidelberg, Germany\\
  $^4$Max Planck Institute for Astronomy, K\"onigstuhl 17, 69117 Heidelberg, Germany\\
  }

\date{Released 2020}

\pubyear{2020}

\begin{document}
\label{firstpage}
\pagerange{\pageref{firstpage}--\pageref{lastpage}}
\maketitle

\begin{abstract}
We present synthetic dust polarization maps of simulated molecular clouds with the goal to systematically explore the origin of the relative orientation of the magnetic field ($\mathbf{B}$) with respect to the cloud sub-structure identified in density ($n$; 3D) and column density ($N$; 2D). The polarization maps are generated with the radiative transfer code POLARIS, which includes self-consistently calculated efficiencies for radiative torque alignment. The molecular clouds are formed in two sets of 3D magneto-hydrodynamical simulations: (i) in colliding flows (CF), and (ii) in the SILCC-Zoom simulations. In 3D, for the CF simulations with an initial field strength below $\sim$5 $\mu$G, $\mathbf{B}$ is oriented either parallel or randomly with respect to the $n$-structures. For CF runs with  stronger initial fields as well as all SILCC-Zoom simulations, which have an initial field strength of 3~$\mu$G, a flip from parallel to perpendicular orientation occurs at high densities of $n_\rmn{trans}$ $\simeq$ 10$^2$ -- 10$^3$ cm$^{-3}$. We suggest that this flip happens if the cloud's mass-to-flux ratio, $\mu$, is close to or below the critical value of 1. This corresponds to a field strength around \mbox{3 -- 5~$\mu$G}, close to the Galactic average. In 2D, we use the method of Projected Rayleigh Statistics (PRS) to study the relative orientation of $\mathbf{B}$. If present, the flip in orientation occurs in the projected maps at $N_\rmn{trans}$ $\simeq$ 10$^{21 - 21.5}$ cm$^{-2}$. This value is similar to the observed transition value from sub- to supercritical magnetic fields in the interstellar medium. However, projection effects can strongly reduce the predictive power of the PRS method: Depending on the considered cloud or line-of-sight, the projected maps of the SILCC-Zoom simulations do not always show the flip, although it is expected given the 3D morphology. Such projection effects can explain the variety of recently observed field configurations, in particular within a single cloud. Finally, we do not find a correlation between the observed orientation of $\mathbf{B}$ and the $N$-PDF.
\end{abstract}

\begin{keywords}
 MHD -- radiative transfer -- methods: numerical -- techniques: polarimetric -- ISM: clouds -- ISM: magnetic fields
\end{keywords}



\section{Introduction}

Magnetic fields appear to play a crucial role in the evolution of gas in disc galaxies, from the diffuse interstellar medium \citep[ISM; see e.g. the reviews by][]{Crutcher12,Beck13} to dense molecular clouds (MCs) and star forming cores \citep[see e.g. the review by][]{Li14}. They can be observed e.g. by polarized radiation emitted from dust grains. In particular recent observations with the BlastPol experiment \citep{Matthews14,Fissel16,Fissel19,Gandilo16,Santos17,Soler17b,Ashton18} and the Planck satellite \citep{PlanckXX,PlanckXXXII,PlanckXXXV} provide more and more dust polarization observations of MCs \citep[see also e.g.][]{Houde04,Dotson10,Li13,Pillai15}.

This dense, molecular part of the ISM is observed to be highly filamentary \citep[see e.g. the review by][]{Andre14}. The impact of magnetic fields on the formation of these filaments and thus finally on the star formation process itself is subject to active investigations \citep[e.g.][]{Goodman92,Goldsmith08,Chapman11,Sugitani11,Li13,Palmeirim13,Malinen16,Panopoulou16, PlanckXXXII,PlanckXXXV,Soler16,Soler17b,Soler19,Jow18,Monsch18,Fissel19}. It has been proposed that the orientation of magnetic field lines with respect to the gas flow and the dense structures/filaments gives insight into whether (i) magnetic fields channel the gas flow along their direction as it would be the case for strong magnetic fields, or (ii) whether the field is dragged along with the flow as it would be the case for weak fields \citep[e.g.][]{Li14}. A general outcome of the aforementioned observations is that there appears to be a progressive change in the relative orientation of the magnetic field from being preferentially parallel to the density structures at low column densities to preferentially perpendicular at high column densities. We emphasise, however, that some recent observations challenge these findings \citep{PlanckXXXV,Soler17b,Soler19,Jow18,Fissel19}, a fact we will investigate in this work.

Results from numerical simulations show that for strong magnetic fields dense structures are mostly perpendicular to the field direction \citep[e.g.][but see also \citealt{Hennebelle19} for a recent review]{Heitsch01,Ostriker01,Li04,Nakamura08,Collins11,Hennebelle13,Soler13,Chen15,Chen20,Li15,Seifried15,Chen16,Zamora17,Mocz18}. This can be attributed to the fact that in ideal magneto-hydrodynamics (MHD) the gas can move freely only along the magnetic field lines whereas the flow perpendicular to it is hampered. The latter is the case when the (turbulent) motions of the gas are sub-Alfv\'enic. Consequently, both gravitating structures like star forming filaments and supersonic shock fronts will be mostly perpendicular to the magnetic field.

\citet{Soler17a} developed a theory to describe the evolution of the angle between the magnetic field and the gas structures. They show that a perpendicular arrangement of magnetic fields and gas structures is a consequence of gravitational collapse or converging flows. \citet{Chen16} argue that the transition from a parallel to perpendicular orientation happens once the flow becomes super-Alfv\*enic due to gravitational collapse. \citet{Soler17a}, however, find that a super-Alfv\*enic flow is necessary but not sufficient for a perpendicular orientation to occur.

Investigating the orientation of magnetic fields in numerical simulations and comparing the results to actual observations is challenging for various technical reasons. First, self-consistent (MHD) simulations have to be performed which capture a wide dynamical range from the larger-scale galactic environment of the clouds down to sub-pc scales. Secondly, the simulations have to include an appropriate treatment for the thermal evolution of both gas and dust. The latter is required for the accurate modelling of dust alignment efficiencies \citep{Lazarian07,Andersson15}, which presents the second challenge. One of the major  obstacles here is the lack of a coherent dust grain alignment theory combining the different alignment processes \citep[see e.g.][for an overview]{Reissl16}. Thirdly, full radiative transfer calculations are required to produce synthetic dust polarization maps. Most of the works presented to date on this topic usually lack at least one of the aforementioned requirements, and thus do not produce \textit{fully} self-consistent dust polarization maps \citep[e.g.][]{Heitsch01b,Ostriker01,Padoan01,Pelkonen07,Pelkonen09,Kataoka12,Soler13,PlanckXX,Chen16,King18,Vaisala18}. In this work we try to overcome these difficulties in the following way:
\begin{itemize}
 \item We use two sets of MC simulation, these are colliding flow simulations \citep{Joshi19} and the SILCC-Zoom simulations \citep{Seifried17} in order to study the relation between polarization observations and the physical (3D) cloud conditions.
 \item  In order to create the polarization maps, we use the freely available dust polarization radiative transfer code POLARIS \citep{Reissl16,Reissl19}, which is able to calculate grain alignment efficiencies and the subsequent radiative transfer in a fully self-consistent manner. The code was already successfully applied in a number of synthetic dust polarization studies from cloud to protostellar disc scales \citep{Reissl17,Seifried19,Valdivia19} as well as the calculation of synthetic synchrotron maps and Zeeman splitting \citep{Reissl18,Reissl19}.
 \item The results of the dust polarization radiative transfer simulations are analysed using the Projected Rayleigh Statistics \citep{Jow18} and are compared to existing observations and are interpreted using the analytical explanation of \citet{Soler17a}.
\end{itemize}

The structure of the paper is as follows: First, we present the initial conditions and various methods used for the MHD simulations and the subsequent radiative transfer with POLARIS (Section~\ref{sec:numericsoverview}). We present our results concerning the colliding flow simulations in Section~\ref{sec:CF} and the SILCC-Zoom simulations in Section~\ref{sec:zoom} and discuss their agreement with the analytical theory of \citet{Soler17a}. In Section~\ref{sec:discussion} we discuss our results in a broader context, before we conclude in Section~\ref{sec:conclusion}.

\section{Numerics, initial conditions and applied methods}
\label{sec:numericsoverview}

In the following we describe the radiative transfer methods, initial conditions and methods used for the colliding flow (CF) simulations and the SILCC-Zoom simulations. As they have been described in detail in previous papers, we only briefly summarise the main points. For more details on the CF simulations we refer to \citet{Joshi19} and for the SILCC-Zoom simulations to \citet{Seifried17,Seifried19}. For the dust polarization radiative transfer we refer to \citet{Reissl16} and \citet{Seifried19}.

\subsection{Numerics}
\label{sec:numerics}

Both the CF and SILCC-Zoom simulations are performed with the adaptive mesh refinement code FLASH 4.3 \citep{Fryxell00,Dubey08}. The CF simulations use a magneto-hydrodynamics solver which guarantees positive entropy and density \citep{Bouchut07,Waagan09}, the SILCC-Zoom simulation an entropy-stable magneto-hydrodynamics solver which guarantees that the smallest possible amount of dissipation is included \citep{Derigs16,Derigs18}. For both types of simulations, we model the chemical evolution of the ISM using a chemical network for H$^+$, H, H$_2$, C$^+$, CO, e$^-$, and O \citep[][but see also \citealt{Walch15} for the implementation in the simulations]{Nelson97,Glover07b,Glover10}.

The simulations follow the thermal evolution of the gas including the most relevant heating and cooling processes. The shielding of the interstellar radiation field \citep[$G_0$ = 1.7 in units of the radiation field of \citet{Habing68} corresponding to the strength determined by][]{Draine78} is calculated according to the surrounding column densities of total gas, H$_2$, and CO via the {\sc TreeRay}/{\sc OpticalDepth} module \citep{Clark12b,Walch15,Wunsch18}. The cosmic ray ionisation rate for atomic hydrogen\footnote{Note that in \citet{Seifried17} we erroneously wrote \mbox{1.3$\times$10$^{-17}$ s$^{-1}$.}} is \mbox{3$\times$10$^{-17}$ s$^{-1}$}. We solve the Poisson equation for self-gravity with a tree based method \citep{Wunsch18}. In addition, for the SILCC-Zoom simulations, we include a background potential from the pre-existing stellar component in the galactic disc, modelled as an isothermal sheet with \mbox{$\Sigma_\mathrm{star}$ = 30 M$_{\sun}$ pc$^{-2}$} and a scale height of \mbox{100 pc} \citep{Walch15,Girichidis16}.

\subsection{Colliding flow simulations}
\label{sec:initial-CF}

The CF simulation domain represents a 128~pc $\times$ 32~pc $\times$ 32~pc rectangular cuboid with inflow boundary conditions in the $x$-direction and periodic boundaries in the $y$- and $z$-direction. The whole domain is initially filled with a warm, uniform density medium with a density of \mbox{$\rho_0$ = 1.67 $\times$ 10$^{-24}$ g cm$^{-3}$} consisting of atomic hydrogen and C$^+$ and an equilibrium temperature of 5540~K. The gas on either side of the $x$ = 0 plane is moving towards the plane with a velocity of $\pm$13.6 km s$^{-1}$ such that the collision occurs immediately upon the start of the simulation. In order to allow turbulent motions to develop, the initial collision plane is not exactly the $x$ = 0 plane but rather represents an irregular interface with the collision taking place at
\begin{equation}
 x = A \left[\rmn{cos}(2 - \tilde y \tilde z) \rmn{cos}(k_y \tilde y) + \rmn{cos}(0.5 - \tilde y \tilde z) \rmn{sin}(k_z \tilde z) \right] \, ,
\end{equation}
with A = 1.6 pc, $k_y$ = 2, $k_z$ = 1, $\tilde y$ = $\pi \cdot y$/(32 pc) and $\tilde z$ = $\pi \cdot z$/(32~pc) \citep[see interface I5 in Fig.~3 of][]{Joshi19}.

The magnetic field is initially homogeneous and parallel to the $x$-axis. In order to test the dependence of our results on the field strength, we perform 5 simulations with magnetic field strengths of $B_{x,0}$ = 1.25, 2.5, 5.0, 7.5, and 10~$\mu$G. Using the collision velocity \mbox{$v$ = 13.6 km s$^{-1}$}, this results in Alfv\'enic Mach numbers
\begin{equation}
 M_\rmn{A} = \frac{v}{B/\sqrt{4 \pi \rho_0}}
 \label{eq:Ma}
\end{equation}
in the moderately sub- to moderately super-alfv\'enic range (see Table~\ref{tab:overview}).

The initial resolution of the simulations is 0.25~pc. During the course of the simulation, we allow for a higher resolution of up to 0.008 pc using a refinement criterion based on the local Jeans length, which must be resolved with at least 8 grid cells in one dimension.

\begin{table}
\caption{Overview of the simulations giving the run name, the initial magnetic field strength and Alfv\'enic Mach number, and the highest resolution reached. Furthermore, we list the reference time $t_0$ to which the times used throughout the paper refer. For the SILCC-Zoom simulations this corresponds to the time at which we start to zoom-in. The second-last column gives the mass-to-flux ratio $\mu$ at \mbox{$t_0$ + 3 Myr} \mbox{(i.e. $t_\rmn{evol}$ = 3 Myr)} and the last column the center of the zoom-in region.}
\begin{tabular}{lcccccc}
  \hline
 run & B$_{x,0}$ & $M_\rmn{A}$ & d$x_\rmn{min}$ & $t_0$ & $\mu$ & center \\
  & ($\mu$G) & & (pc)  &  (Myr) & & (pc) \\
 \hline 
 CF-B1.25 & 1.25 & 4.0 & 0.008 & 16.0 & 4.3 & --- \\
 CF-B2.5 & 2.5 & 2.5 & 0.008 & 16.0 & 2.2 & --- \\
 CF-B5 & 5.0 & 1.2 & 0.008 & 16.0 & 1.1 & --- \\
 CF-B7.5 & 7.5 & 0.8 & 0.008 & 16.0 & 0.72 & --- \\
 CF-B10 & 10 & 0.6 & 0.008 & 16.0 & 0.54 & --- \\
 \hline
 SILCC-MC1 & 3.0 & 1.8 & 0.12 & 16.0 & 2.2 & (-84, 100, 0) \\
 SILCC-MC2 & 3.0 & 1.8 & 0.12 & 16.0 & 2.8 & (126, -117, 0) \\
 SILCC-MC3 & 3.0 & 1.8 & 0.12 & 16.0 & 2.1 & (-125, -104, 0) \\
 SILCC-MC4 & 3.0 & 1.8 & 0.12 & 11.6 & 2.1 & (-97, 130, 0) \\
 SILCC-MC5 & 3.0 & 1.8 & 0.12 & 11.6 & 2.3 & (3, 16, 0) \\
 SILCC-MC6 & 3.0 & 1.8 & 0.12 & 16.0 & 1.4 & (62, 175, 0) \\
 \hline
 \end{tabular}
 \label{tab:overview}
\end{table}

\subsection{SILCC-Zoom simulations}
\label{sec:initial-zoom}

In the following we briefly describe the SILCC-Zoom setup, for more details we refer to \citet{Seifried17}. The simulation domain represents a 500~pc $\times$ 500~pc $\times$ $\pm$5~kpc section of a galactic disc with an initial resolution of 3.9 pc. The initial gas distribution follows a Gaussian profile
\begin{equation}
 \rho(z) = \rho_0 \times \textrm{exp}\left[ - \frac{1}{2} \left( \frac{z}{h_z} \right)^2 \right] \, ,
 \label{eq:rhosilcc}
\end{equation}
with $h_z$ = 30 pc and $\rho_0 = 9 \times 10^{-24}$ g cm$^{-3}$. This results in a gas surface density of \mbox{$\Sigma_\rmn{gas}$ = 10 M$_{\sun}$ pc$^{-2}$}, similar to the solar neighbourhood. The initial magnetic field is given by
\begin{equation}
 B_{x} = B_{x,0} \sqrt{\rho(z)/\rho_0} \; , B_y = 0 \; , B_z = 0 \, ,
 \label{eq:bsilcc}
\end{equation}
with the magnetic field in the midplane being set to \mbox{$B_{x,0}$ = 3 $\mu$G} following recent observations \citep[e.g.][]{Beck13}. Assuming a typical turbulent velocity dispersion of \mbox{5 km s$^{-1}$} \citep[figure~5 in][]{Seifried17}, we obtain slightly super-Alfv\'enic turbulent motions (Table~\ref{tab:overview}).

From the start we inject supernovae (SNe) up to a certain time $t_0$ with a constant rate of 15 SNe Myr$^{-1}$. The time $t_0$ was chosen such that the SNe can generate sufficient turbulent motions in the simulation domain \citep[see Section~2 in][]{Seifried18}. We use mixed SN driving, which allows us to obtain a realistic distribution of the multiphase ISM as initial conditions for the subsequent zoom-in procedure \citep{Walch15,Girichidis16}. If the Sedov-Taylor radius is resolved with at least 4 cells, we inject 10$^{51}$~erg per SN in form of thermal energy. Otherwise, the gas inside the injection region is heated to $10^4$~K and momentum corresponding to the end of the pressure-driven snowplough phase is injected \citep[see][for details]{Gatto17}.

At $t_0$ we stop the injection of further SNe and choose six different cuboid-like regions centered in the midplane of the disc. In order to follow the evolution of the clouds forming in these six regions -- henceforth denoted as MC1 to MC6 -- we then progressively increase the spatial resolution inside these regions from 3.9~pc to 0.12~pc \citep[see Table 2 in][]{Seifried17}, refining based on the Jeans length and variations in the gas density. Afterwards we keep the highest resolution of 0.12 pc in the zoom-in regions, and the lower resolution of 3.9 pc outside. For all six clouds the corresponding $t_0$ and the centers of the zoom-in regions are listed in Table~\ref{tab:overview}.

\subsection{POLARIS and radiative transfer}
\label{sec:polaris}

The radiative transfer (RT) calculations are performed with the freely available RT code POLARIS\footnote{http://www1.astrophysik.uni-kiel.de/$\sim$polaris} \citep{Reissl16,Reissl19}. POLARIS is a 3D line and dust continuum Monte-Carlo code allowing to solve the RT problem including dust polarization, which we have already successfully applied before \citep{Reissl17,Seifried19}.

We apply the radiative torque (RAT) alignment theory \citep{Dolginov76,Draine96,Draine97,Bethell07,Lazarian07,Hoang08,Andersson15}. In short, using RAT, POLARIS determines the Stokes parameters by calculating the size-dependent alignment of dust grains with the magnetic field. The dust temperature is provided by the MHD simulations. Dust grains smaller than the threshold size $a_\rmn{alig}$ are not aligned with the magnetic field as the spinning-up due to the incident radiation is smaller than the randomizing effect of collisions with gas particles. The upper threshold, up to which grains are still aligned, is given by the Larmor limit, $a_l$, which is typically of the order of, or larger, than the maximum grain size assumed here (see below). We use the ISRF of \citet{Mathis77} scaled up by a factor of 1.47 such that its strength corresponds to that determined by \citet{Draine78}.

As we focus on MCs in a Galactic environment, we apply a dust model consisting of 37.5\% graphite and 62.5\% amorphous silicate grains \citep{Mathis77}. The dust density is obtained from the gas density assuming a spatially constant dust-to-gas mass ratio of 1\%. We assume a grain size distribution of $n(a) \propto a^{-3.5}$ with the canonical values of the lower and upper cut-off radius of $a_{\rm min}$ = 5 nm and \mbox{$a_{\rm max}$ = 2 $\mu$m}, respectively, the latter accounting for a moderate grain growth in the dense ISM. The shape of a single dust grain is fractal in nature. However, we apply an oblate shape with an aspect ratio of $s = 0.5$, a valid approximation for an averaged ensemble of dust grains \citep{Hildebrand95,Draine17}. We pre-calculate individual cross sections for 160 size bins and 104 wavelength bins \citep[see][for details]{Reissl17} with the scattering code DDSCAT \citep{Draine13}. Optical properties of the different materials are taken from tabulated data of \cite{Lee85} and \citet{Laor93}.

Here, we focus on RT calculations for a wavelength of \mbox{$\lambda$ = 1.3 mm}, which is close to the wavelength of the CO(2-1) transition. We emphasise that choosing e.g. $\lambda$ around 850 $\mu$m as in the Planck observations would give qualitatively and quantitatively very similar results, as the polarization maps show only very little difference of the order of 1$^\circ$ for the various wavelengths \citep{Seifried19}. The spatial resolution is identical to the highest resolution of the corresponding MHD simulation, i.e. 0.008~pc for the CF runs and 0.12~pc for the SILCC-Zoom runs. In order to mimic the observation of MCs forming in isolation and to avoid confusion of the polarization signal along the line-of-sight, we perform the RT calculations for a cubic sub-region of each simulation domain. For the CF runs, we pick a \mbox{32~pc $\times$ 32~pc $\times$ 32~pc} region centered in the middle of the simulation domain covering the entire collision interface. For the SILCC-Zoom runs we take a \mbox{125~pc $\times$ 125~pc $\times$ 125~pc} region centered on the midpoint of the corresponding zoom-in region (see Table~\ref{tab:overview}).
	
From the RT calculations we obtain the Stokes parameters $I$, $Q$, and $U$, where $I$ is the total intensity, and $Q$ and $U$ quantify the linear polarization of the observed radiation \citep[see][for details]{Reissl16,Reissl19}. The polarization angle, $\phi_\rmn{Pol}$, is calculated as
\begin{equation}
 \phi_\rmn{Pol} = \frac{1}{2}\rmn{arctan}(U,Q) \,
 \label{eq:phiPol}
\end{equation}
and the polarization degree is given by 
\begin{equation}
 p = \frac{\sqrt{Q^2 + U^2}}{I} \, .
 \label{eq:p}
\end{equation}

\subsection{The Projected Rayleigh Statistic}
\label{sec:prs}

The tools of Rayleigh Statistic were first applied to astrophysical problems by \citet{Jow18}. As discussed there, the Rayleigh Statistic tests whether in 2D for a set of $n$ independent angles $\theta_i$ in the range $[0,2\pi]$ the angles are uniformly distributed by calculating
\begin{equation}
 Z = \frac{ (\Sigma_i^{n} \rmn{cos}\theta_i)^2 + (\Sigma_i^{n} \rmn{sin}\theta_i)^2 }{n} \, .
\end{equation}
This equation is identical to a random walk in 2D with $Z$ being the displacement from the origin if steps of unit length in the direction of $\theta_i$ are taken.

In order to test the relative orientation of the magnetic field direction and density structures, we take $\theta$ = 2 $\phi$, where $\phi$ is the relative orientation angle between the plane-of-sky projected magnetic field $\mathbf{B_\rmn{POS}}$, inferred from the polarization direction by rotating it by 90$^\circ$, and the tangent to the column density ($N$) isocontour \citep{Soler17b}. This is equivalent to the angle between the observed polarization direction $\mathbf{E}$ and the gradient of the column density, $\nabla N$, which allows us to calculate the angle as
\begin{equation}
 \phi = \rmn{arctan}\left( |\nabla N \times \mathbf{E}|, \nabla N \cdot \mathbf{E} \right) \, .
 \label{eq:phi}
\end{equation}

We correct for a possible oversampling of our data by checking against 100 realizations of a randomly distributed $\nabla N$ map as discussed in detail in \citet[][see their Eqs.~11 and~12]{Fissel19}. This takes into account that in our synthetic dust polarization maps neighbouring pixels (in particular in the less resolved, lower-density regimes) are not statistically independent. This effectively reduces the total number of pixels from $n$ to $n_\rmn{ind}$ where $n_\rmn{ind}$ ($<$ $n$) is the number of independent data samples.

As discussed by \citet{Jow18}, the Projected Rayleigh Statistic (PRS) denoted with the symbol $Z_x$ can be used to test whether a preferred parallel or perpendicular orientation is present:
\begin{equation}
 Z_x = \frac{ \Sigma_i^{n_\rmn{ind}} \rmn{cos}\theta_i }{\sqrt{n_\rmn{ind}/2}} \, .
\end{equation}
If the observed magnetic field $\mathbf{B_\rmn{POS}}$ is parallel to the iso-$N$ contour, then cos$\theta_i$ = 1; if the two directions are perpendicular, then cos$\theta_i$ = -1. Hence, measurements of $Z_x$ $\gg$ 0 indicate preferentially parallel orientation, whereas $Z_x$ $\ll$ 0 indicates magnetic fields perpendicular to the $N$ isocontours. For $Z_x$ $\simeq$ 0, no preferred direction is present. Finally, in order to test the statistical significance of the orientation, we compare $Z_x$ against its variance \citep{Jow18}
\begin{equation}
 \sigma^2_{Z_x} = \frac{2 \Sigma_i^{n_\rmn{ind}} (\rmn{cos} \theta_i)^2 - (Z_x)^2 }{n_\rmn{ind}} \, .
\end{equation}

\section{Results of the CF simulations}
\label{sec:CF}

\begin{figure*}
 \includegraphics[width=\textwidth]{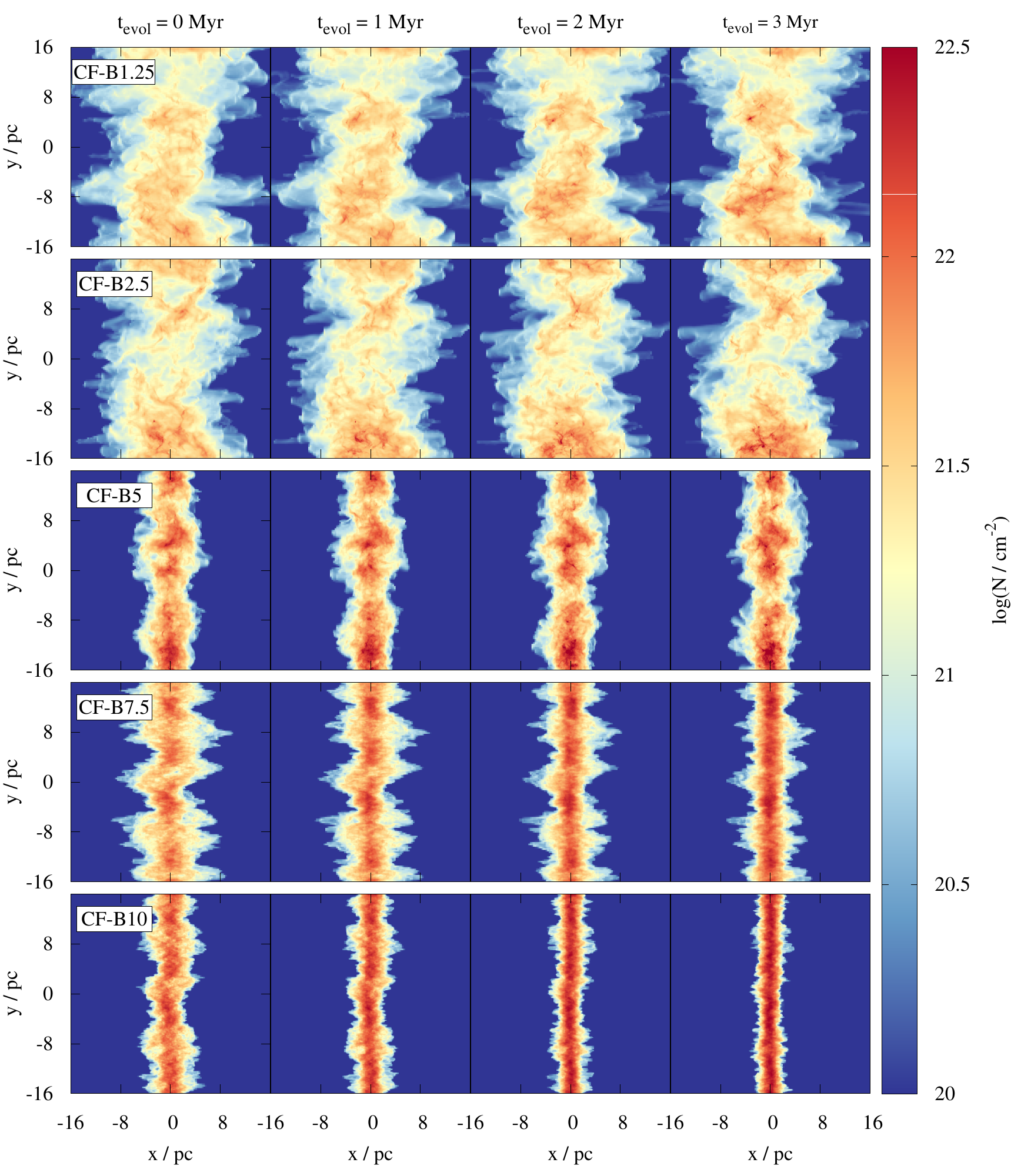}
 \caption{Time evolution (from left to right) of the column density of the CF runs with increasing magnetic field strength (from top to bottom). The higher the magnetic field strength, the more confined is the cloud to the collision interface around $x$ = 0. In addition, stronger magnetic fields suppress structure formation perpendicular to the original field direction. Note that the panels only show the central part of the simulation domain.}
 \label{fig:CD_CF}
\end{figure*}

\begin{figure*}
 \includegraphics[height=0.44\textwidth]{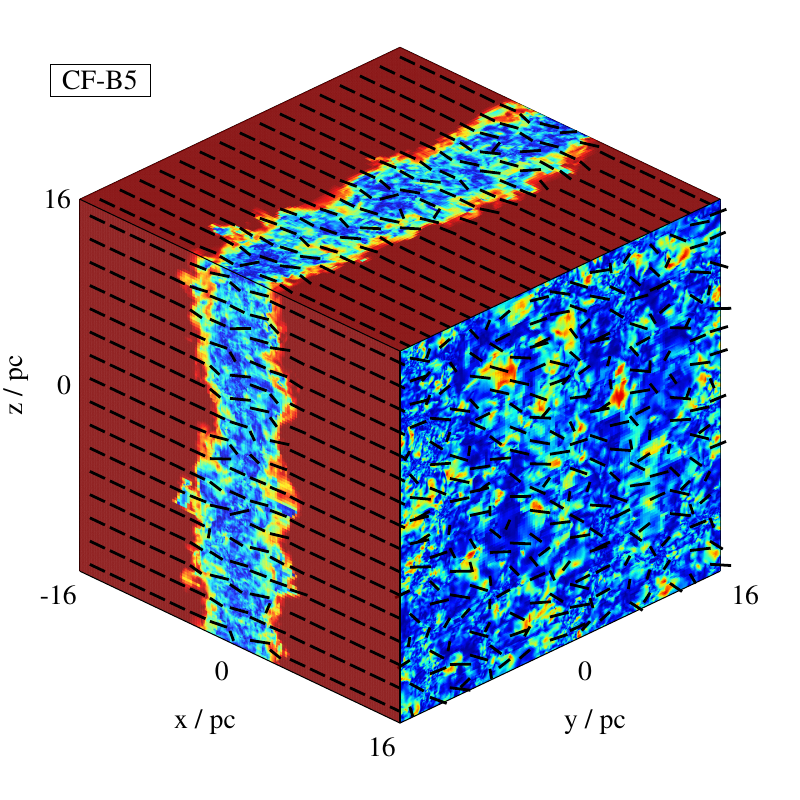}
 \includegraphics[height=0.44\textwidth]{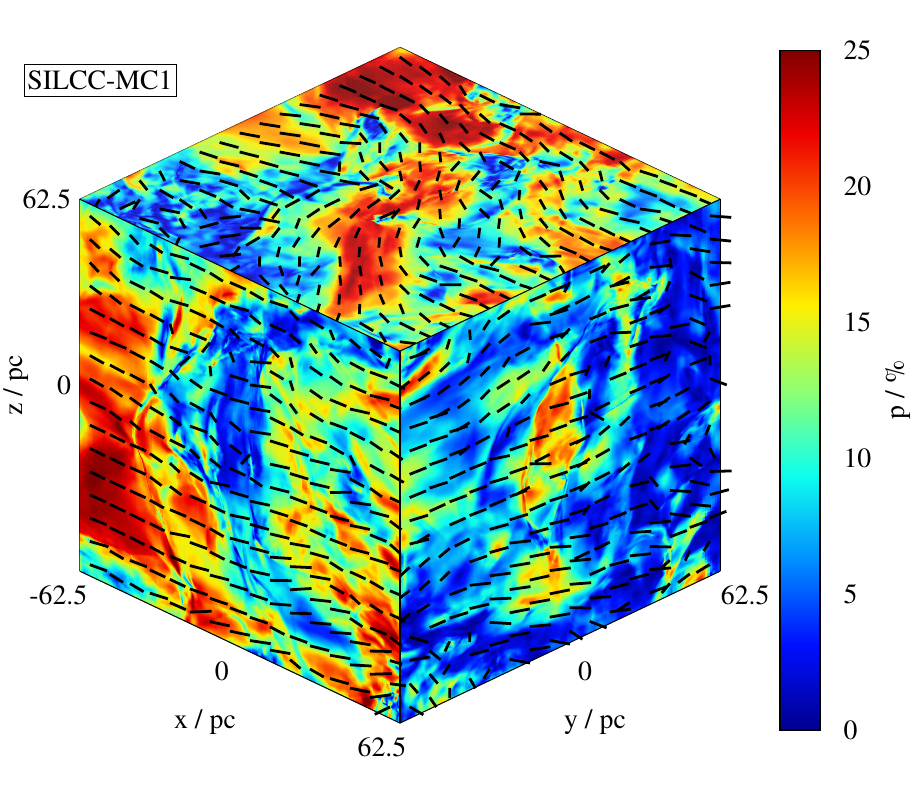}
 \caption{Synthetic polarization maps obtained with POLARIS of run CF-B5 (left) and SILCC-MC1 (right) at $t_\rmn{evol}$ = 3 Myr. The figure shows the polarization degree from 3 orthogonal directions (colour coded) and the polarization direction rotated by 90$^\circ$ (black bars) to view the inferred magnetic direction.}
 \label{fig:pol_maps}
\end{figure*}

For the CF simulations we focus on the results from \mbox{$t_0$ = 16 Myr} onwards. By inspecting the simulations, this is the time when sufficient mass ($\sim$ 10$^{4}$ M$_{\sun}$) has accumulated at the collision interface and gravitational collapse accompanied by the formation of dense molecular gas sets in. We note that throughout the paper we refer to the time elapsed since $t_0$ as $t_\rmn{evol}$ = $t$ - $t_0$. We note that we first investigate the relative orientation between the observed magnetic field and the column density (Section~\ref{sec:prs_CF}), and then link its result to the underlying 3D structure (Section~\ref{sec:a23_CF}).

In Fig.~\ref{fig:CD_CF} we show the time evolution of the column density of the five CF runs from $t_\rmn{evol}$ = 0 -- 3 Myr. For all runs the accumulation of mass in the central region can be observed. For the runs with the highest magnetic field strengths, CF-B7.5 and CF-B10, the dense regions appear to contract along the $x$-direction and are confined to $\sim$ $\pm$5 pc around $x$ = 0. For the remaining runs this contraction is less clear. In particular for the low magnetic field runs the collision region remains rather widespread with a typical extent of 15 -- 20~pc. This indicates that for CF-B7.5 and CF-B10 the magnetic field is strong enough to guide the gas streams, to promote a collapse along its original direction, and to prevent structure formation by turbulent motions perpendicular to it \citep[see also][]{Heitsch09,Zamora18,Iwasaki19}. For the remaining runs the gas is able to form significant structures also perpendicular to the original field direction. For a more detailed discussion on the dynamical and chemical evolution of the clouds we refer to Weis et al. (in prep.).

In the left panel of Fig.~\ref{fig:pol_maps} we show the map of the polarization degree and direction for run CF-B5 at $t_\rmn{evol}$ = 3 Myr. In the collision region (parallel to the $y$-$z$-plane) the polarization degree is typically of a few 1 to 10\%. Only in the inflowing low-column density medium, where the field is still well ordered and resembles the initial setup, high polarization degrees ($\geq$ 25\%) are obtained. We note, however, that these would probably not be accessible in actual observations due to a low signal-to-noise ratio \citep{Seifried19}. Qualitatively similar results are also found for the other runs and times, although the drop in polarization degree in the collision region for the runs CF-B7.5 and CF-B10 is less pronounced.

\subsection{2D: The relative orientation between the observed magnetic field and $\nabla N$}
\label{sec:prs_CF}

\begin{figure*}
 \includegraphics[width=\textwidth]{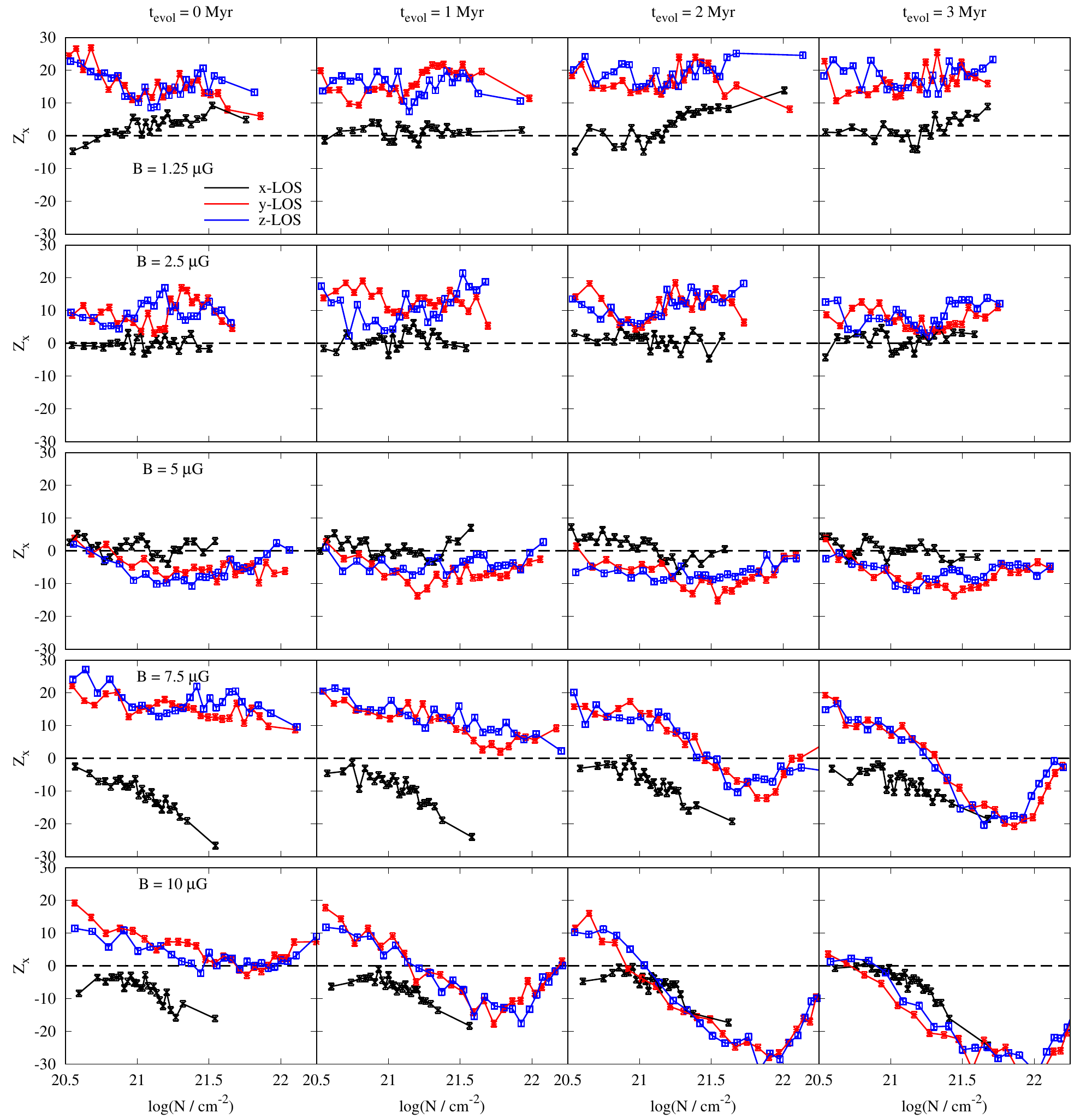}
 \caption{Time evolution (from left to right) of the PRS, $Z_x$, of the CF runs with increasing magnetic field strength (from top to bottom) for three different LOS. A preferentially perpendicular orientation ($Z_x <$ 0) at high column densities is only present for runs with field strength $\geq$ 5 $\mu$G. In addition, for these runs the perpendicular orientation becomes more pronounced at later times. For the weak field runs, the magnetic field remains mostly parallel to the density structures or shows no preferred direction. The uncertainty $\sigma_{Z_x}$ of each bin (shown by horizontal bars) is rather minor.}
 \label{fig:PRS_CF}
\end{figure*}

In Fig.~\ref{fig:PRS_CF} we show the PRS, $Z_x$, inferred from the polarization maps (Fig.~\ref{fig:pol_maps}) of the five CF runs for three lines-of-sights (LOS) as a function of time, i.e. for the same snapshots as shown in Fig.~\ref{fig:CD_CF}. For this purpose, we use the freely available tool magnetar\footnote{https://github.com/solerjuan/magnetar} \citep{Soler13}. Before evaluating the PRS, we smooth the polarization and column density maps with a Gaussian kernel with a size of 5 pixels, in order to average over regions in which the resolution of the MHD simulation was lower than the pixel size of 0.008~pc used in the polarization maps. We have chosen the bins in $N$ such that each contains the same number of pixels \citep{Soler13}. Furthermore, as stated in Section~\ref{sec:prs}, we have corrected for a potential oversampling of the data. Overall, the typical uncertainty $\sigma_{Z_x}$ is close to 1 \citep{Fissel19} and thus, it is rather small compared to $Z_x$.

The most striking feature is that a preferentially perpendicular orientation of magnetic fields and column density structures ($Z_x <$ 0) is only obtained for the runs with fields strengths \mbox{$B_{x,0}$ $\geq$ 5 $\mu$G}, although for run CF-B5 the relative orientation shows partly no preferred direction within the uncertainty. For the runs CF-B7.5 and CF-B10 the configuration flips from a parallel to a perpendicular configuration around column densities of \mbox{$N_\rmn{trans}$ $\simeq$ 10$^{21 - 21.5}$ cm$^{-2}$}. This value is at the lower end of the distribution seen in recent observations~\citep{PlanckXXXV,Jow18,Soler17b,Soler19}. Moreover, the value of $N_\rmn{trans}$ agrees very well with the value of \citet{Crutcher12}, at which the transition from a sub- to a supercritical magnetic field in the ISM occurs.

Interestingly, for the runs CF-B7.5 and CF-B10 we find a clear time evolution, which is not the case for the other runs. As time progresses the relative orientation becomes increasingly more perpendicular for high column densities. Also the values of $N_\rmn{trans}$ decrease over time. In addition, by comparing the runs CF-B7.5 and CF-B10, we find that the transition from parallel to perpendicular orientation appears to occur at lower column densities for higher magnetic field strengths. Some of the curves are even located completely below $Z_x$ = 0 \citep[see e.g. Fig. A.2 in][for an observational counterpart]{Soler17b}.

For the runs CF-B1.25 and CF-B2.5, the PRS shows a preferentially parallel orientation for the $y$- and $z$-direction, whereas for the projection along the $x$-direction, i.e. along the original field direction, there is no preferred direction of the magnetic field recognisable. For this latter projection, we partly see an increase of $Z_x$ with increasing column density for run CF-B1.25. We attribute this to the fact that for the lower column densities (corresponding to the inflowing material) the orientation is random, whereas for the  highest column densities, i.e. the collision interface, a parallel orientation, similar to the other the directions, tries to establish. Overall, however, we do no see a clear time evolution for the low-magnetic field runs.

\subsubsection{Link to the column density evolution}

The general shape of the PRS can be directly linked to the evolution of the column density (Fig.~\ref{fig:CD_CF}): For the low magnetic field runs, the column density structure remains rather unchanged over time and shows elongated structures along the $x$-direction. Consequently, as the magnetic field is mainly along the $x$-direction, for both the $y$- and $z$-projection (red and blue curves in Fig.~\ref{fig:PRS_CF}) the PRS indicates a rather parallel field-density configuration.

For the higher magnetic field runs, however, the column density structure is more confined to the collision interface around $x$ = 0 and thus shows an overall elongated structure along the $y$- and $z$-direction. Together with a magnetic field along the $x$-direction, this results in a perpendicular configuration. In addition, the observed contraction of the structures along the $x$-direction and the increase in the maximum column density over time is also visible in the PRS: the more contracted, the clearer the perpendicular orientation.

\subsection{3D: The relative orientation of the magnetic field and $n$}
\label{sec:a23_CF}

\begin{figure}
\includegraphics[width=\linewidth]{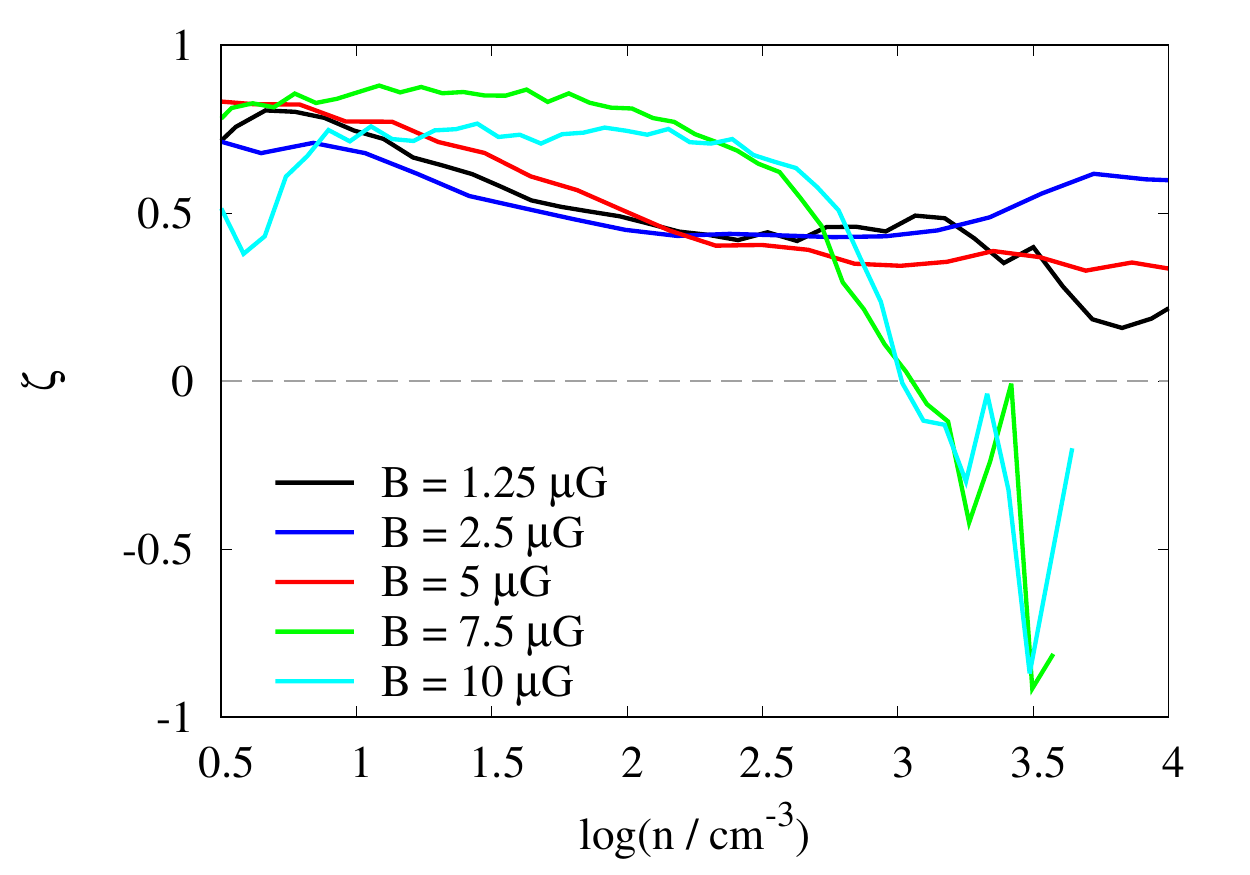}
\caption{Dependence of $\zeta$ on the number density for the CF runs at \mbox{$t_\rmn{evol}$ = 2 Myr}. Only for the highly magnetised runs CF-B7.5 and CF-B10, negative values, indicating a perpendicular orientation of the magnetic field and the densest structures, are reached. This matches well the results of the 2D analysis (Fig.~\ref{fig:PRS_CF}).}
\label{fig:zeta_CF}
\end{figure}

\begin{figure}
 \includegraphics[width=0.9\linewidth]{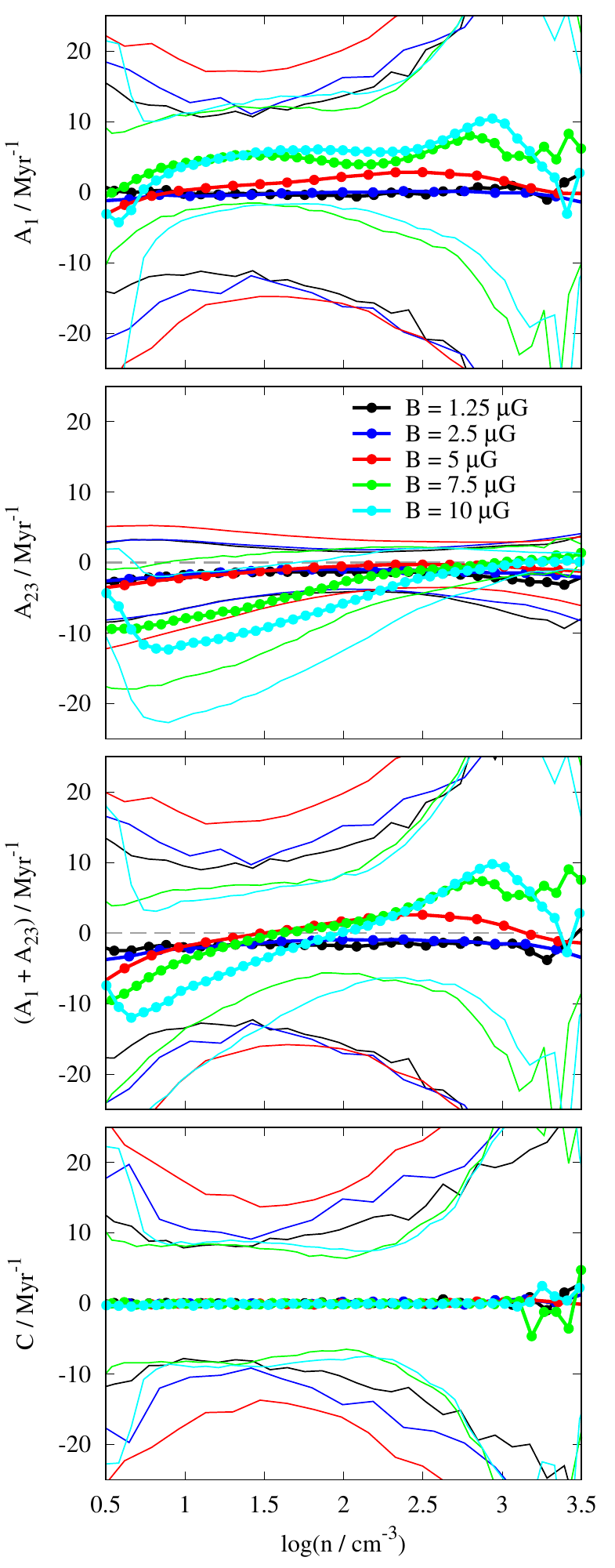}
 \caption{Density dependence of $A_1$, $A_{23}$, $A_1$ + $A_{23}$ and $C$ (from top to bottom) for the five CF runs at $t_\rmn{evol}$ = 2 Myr. The thick lines with dots show the mean value for a given density, the thin lines the 1$\sigma$ interval. Overall, the observed relative orientation in Fig.~\ref{fig:PRS_CF} can be explained by the sum of $A_1$ and $A_{23}$: a positive sum results in perpendicular orientation ($\zeta < 0$ and $Z_x < 0$), a negative sum in parallel orientation ($\zeta > 0$ and $Z_x > 0$). Note, however, that also the large spread of e.g. $C$ can contribute to a perpendicular orientation (Appendix~\ref{sec:appendixA}).}
 \label{fig:a23_CF}
\end{figure}

In order to relate the results found for the 2D polarization maps to the actual conditions in the clouds, we next investigate the relative orientation of the magnetic field and the number density $n$ in 3D. For this purpose we consider the relative orientation angle between the (3D) magnetic field and the \textit{gradient} of the density, \mbox{$\varphi$ = $\measuredangle (\mathbf{B}, \nabla n$)} (in contrast to $\phi$ used for the PRS in 2D, which gives the angle between the projected magnetic field and the \textit{isocontour} of the column density), which we calculate via
\begin{equation}
\rmn{cos} \,\varphi = \frac{\nabla n \cdot\mathbf{B}}{|\nabla n|\,|\mathbf{B}|} \, .
\label{eq:varphi}
\end{equation}
This implies that \mbox{cos $\varphi$ = $\pm$ 1} for $\bf{B}$ perpendicular to the iso-density contours, which are by definition normal to $\nabla n$, and \mbox{cos $\varphi$ = 0} for $\bf{B}$ parallel to the iso-density contours.

The distribution of the angles between randomly oriented pairs of vectors in 3D is only flat in terms of their cosine, as a consequence of the coverage of the solid angle. Hence, we investigate the distribution of cos~$\varphi$, which we evaluate using the relative orientation parameter defined in \citet{Soler13}\footnote{We checked that using $\varphi$ instead of cos~$\varphi$ would not significantly change the qualitative behaviour of our findings.}:
\begin{equation}
 \zeta = \frac{A_c - A_e}{A_c + A_e} \, .
\end{equation}
Here, $A_c$ is the volume, i.e. all cells, in a given density interval for which \mbox{|cos $\varphi$| $<$ 0.25} and $A_e$ the volume for which \mbox{|cos $\varphi$| $>$ 0.75}. In 3D \mbox{$\zeta$ $>$ 0} describes a parallel orientation of the magnetic field with respect to the density structures and \mbox{$\zeta$ $<$ 0} a perpendicular orientation. We note that we here revert to the usage of $\zeta$ as there is no equivalent of the PRS in 3D, where the topology of the space of orientations is different to 2D. This does not imply any loss of generality in our analysis and its outcome is directly comparable to that of other works in the literature \citep[e.g.][]{Soler13,Soler17a}.

In Fig.~\ref{fig:zeta_CF} we show the dependence of $\zeta$ on the number density for \mbox{$t_\rmn{evol}$ = 2 Myr}. Overall, the results of the 3D analysis match those of the 2D analysis (Fig.~\ref{fig:PRS_CF}). Only for the highly magnetised runs CF-B7.5 and CF-B10, $\zeta$ reaches negative values at high densities (\mbox{$n_\rmn{trans}$ $\simeq$ 10$^3$ cm$^{-3}$}), indicating a perpendicular orientation of the magnetic field and the densest structures. For the other runs, $\zeta$ stays above zero in agreement with the parallel and random orientation found in the 2D polarization maps. We note that \citet{Chen16}, who study the ISM in a very small box of 1 pc$^3$, suggest that $\zeta$ drops below zero once the turbulence becomes super-Alfv\'enic. Our analysis, however, reveals that $\zeta$ becomes negative only for $M_A$ $\gg$ 1 (not shown) in agreement with the findings of \citet[][see their figure~6]{Soler17a}. Furthermore, the initial values of $M_A$ (Table~\ref{tab:overview}) show super-alfv\'enic motions for the runs with a parallel relative orientation and vice versa.

\subsubsection{Comparison with \citet{Soler17a}}

In order to investigate the origin of the relative orientation of magnetic fields and density structures in 3D, we apply the theory developed by \citet{Soler17a} to our data, which starts from the continuity equation
\begin{equation}
 \frac{\rmn{d} \, \textrm{log} \rho}{\rmn{d} t} = -\partial_i v_i \, .
\end{equation}
Here, d/d$t$ denotes the Lagrangian time derivative, $v_i$ is the $i$-th component of the gas velocity, $\partial_i$ = $\partial/\partial x_i$ and we use Einstein's sum convention. The authors derive an evolution equation for cos~$\varphi$ (Eq.~\ref{eq:varphi}) which reads:
\begin{equation}
 \frac{\rmn{d(cos}\, \varphi)}{\rmn{d}t} = C +\left[ A_1 + A_{23} \right] \rmn{cos}\, \varphi
 \label{EQ:COSPHI}
\end{equation}
using the definitions
\begin{equation}
 C \equiv -\frac{\partial_i ( \partial_j v_j )}{(R_k R_k)^{1/2}} b_i \, ,
 \label{eq:c}
\end{equation}
\begin{equation}
 A_1 \equiv \frac{\partial_i ( \partial_j v_j )}{(R_k R_k)^{1/2}} r_i \, ,
 \label{eq:a1}
\end{equation}
and
\begin{equation}
 A_{23} \equiv \partial_i v_j \left[r_i r_j - b_i b_j \right] \, .
 \label{eq:a23}
\end{equation}
Here, $b_i$ and $r_i$ are the components of the unity vector pointing in the direction of the magnetic field and of \mbox{$R_i \equiv \partial_i$ log $\rho$}, i.e.
\begin{equation}
b_{i} \equiv \frac{B_{i}}{(B_{k}B_{k})^{1/2}} = \frac{B_{i}}{|\mathbf{B}|} \; \rmn{and} \; r_{i} \equiv \frac{R_{i}}{(R_{k}R_{k})^{1/2}} = \frac{R_{i}}{|\mathbf{R}|} \, .
\label{eq:bi}
\end{equation}

There are a three aspects in Eq.~\ref{EQ:COSPHI} being worth mentioning. First, at \mbox{cos~$\varphi$ = $\pm$1}, the right-hand-side of Eq.~\ref{EQ:COSPHI} becomes zero as \mbox{$r_i$ = $\pm b_i$} (Eq.~\ref{eq:varphi}). Hence, \mbox{cos~$\varphi$ = $\pm$1} is an equilibrium point, at which the magnetic field is perpendicular to the density structures. Second, assuming that C is negligible compared to \mbox{$A_1 + A_{23}$}, also the point \mbox{cos~$\varphi$ = 0} is an equilibrium point. Here, however, the magnetic field is parallel to the density structures. Third, with C being very small, over time cos~$\varphi$ tends towards $\pm$1, when  \mbox{$A_1 + A_{23} >$ 0}, whereas it tends towards 0, when \mbox{$A_1 + A_{23} <$ 0.}

We now calculate $A_1$, $A_{23}$, their sum and $C$ for the CF runs at $t_\rmn{evol}$ = 2~Myr from the 3D simulation data and show the results in Fig.~\ref{fig:a23_CF}. We find that the mean values (thick lines with dots) of $C$ are around 0.1 Myr$^{-1}$, which is about a factor of 10 smaller than the typical mean values of $A_1$ and $A_{23}$, which are of the order of a few 1~Myr$^{-1}$. For this reason, for the moment we can neglect $C$ in our consideration and focus on the interpretation of the mean values only (but see also Section~\ref{sec:a23_zoom} and Appendix~\ref{sec:appendixA}). The variables $A_1$ and $A_{23}$ show both negative and positive mean values with absolute values up to \mbox{$\sim$ 10 Myr$^{-1}$} which overall tend to increase with increasing magnetic field strength.

For the runs CF-B1.25 and CF-B2.5 (black and blue curves), $\left\langle A_1 \right\rangle$ is close to zero with typical values around a few \mbox{$\pm$ 0.1 Myr$^{-1}$}, whereas $\left\langle A_{23} \right\rangle$ remains negative with values around a few times \mbox{-1 Myr$^{-1}$}. This is in excellent agreement with the preferentially parallel orientation of the magnetic field and the density structures shown in the Fig.'s~\ref{fig:PRS_CF} and~\ref{fig:zeta_CF} ($Z_x > 0$ and $\zeta > 0$).

For run CF-B5 (red curves), $\left\langle A_1 \right\rangle$ is mostly positive with values around a few 1 Myr$^{-1}$, which -- at lower densities -- is balanced by negative values of $\left\langle A_{23} \right\rangle$. However, in particular towards higher densities, $\left\langle A_1+ A_{23}\right\rangle$ reaches positives values, indicating a perpendicular orientation as it is indeed the case for $t_\rmn{evol}$ $\geq$ 2 Myr (Fig.~\ref{fig:PRS_CF}).

For the runs CF-B7.5 and CF-B10 (green and cyan curves), $\left\langle A_1 \right\rangle$ and $\left\langle A_{23} \right\rangle$ reach larger values than for the other runs. At low densities $\left\langle A_1 + A_{23} \right\rangle$ is dominated by the negative values of $\left\langle A_{23} \right\rangle$ of $\sim$ -10~Myr$^{-1}$. Towards higher densities, however, $\left\langle A_{23} \right\rangle$ increases and $\left\langle A_1 + A_{23} \right\rangle$ becomes positive in agreement with the clear perpendicular relative orientation of the magnetic field and the density shown in the two bottom rows of the Fig.'s~\ref{fig:PRS_CF} and~\ref{fig:zeta_CF} (\mbox{$Z_x < 0$} and \mbox{$\zeta < 0$}).

In summary, the observed relative orientation in 2D and 3D can be well explained by $\left\langle A_1 + A_{23} \right\rangle$, where both $A_1$ and $A_{23}$ appear to contribute in a comparable manner. The mean value of C is smaller by a factor of $\sim$ 10 -- 100 and thus less likely to contribute. The 1$\sigma$ interval (thin lines in Fig.~\ref{fig:a23_CF}), however, is well comparable to that of $A_{1}$ and $A_{23}$. We emphasise that such a wide distribution of $C$ can also contribute to a preferentially perpendicular orientation (see Section~\ref{sec:a23_zoom} and Appendix~\ref{sec:appendixA}), even if $A_1$ and $A_{23}$ are on average slightly negative. Furthermore, the importance of $A_{23}$ is in agreement with the findings of \citet{Soler17a}. As explained by the authors, $A_{23}$ describes the complex interplay of compressive velocity modes and the magnetic field (see their sections~3.1.1 and~3.1.3). However, contrary to \citet{Soler17a}, where $\left\langle A_1 \right\rangle$ and $\left\langle C \right\rangle$ are about 10$^4$ times smaller than $\left\langle A_{23} \right\rangle$ and thus negligible, here we find that $\left\langle A_1 \right\rangle$ also contributes significantly and that $\left\langle C \right\rangle$ is smaller by a factor of $\sim$ 10 -- 100 only. We speculate that these differences might be due to the different physical setup the authors use, which is why we do not follow this further here.

\section{Results of the SILCC-Zoom simulations}
\label{sec:zoom}

\begin{figure*}
\flushleft
 \includegraphics[height=0.3\textwidth]{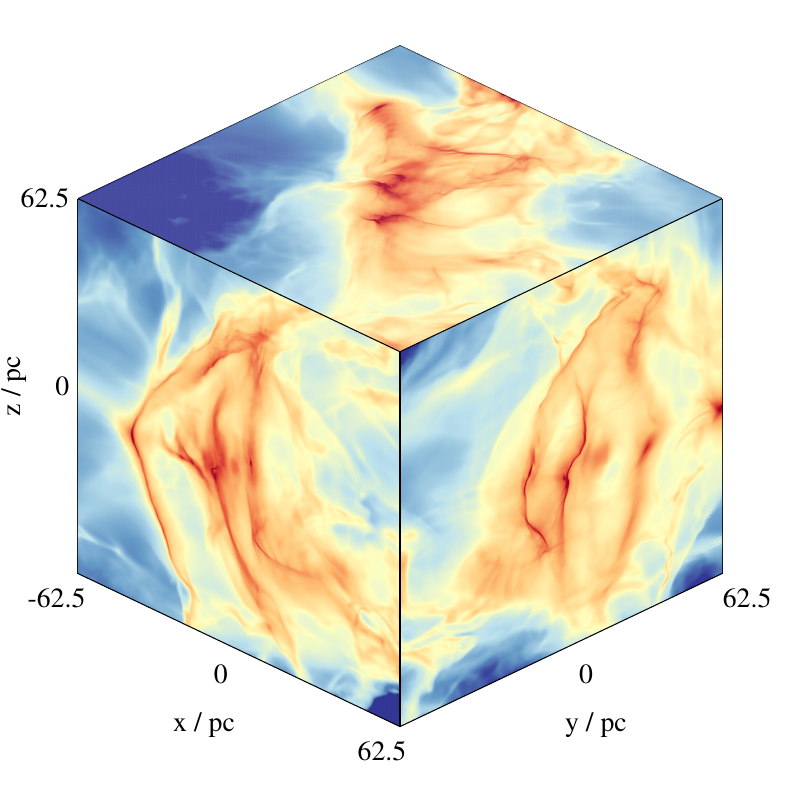}
 \includegraphics[height=0.3\textwidth]{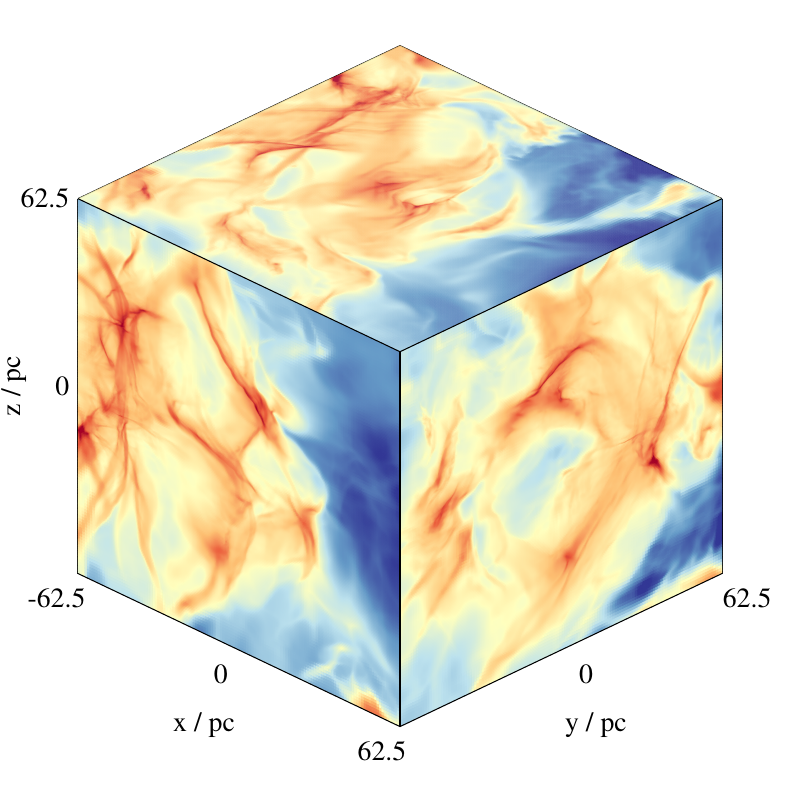}
 \includegraphics[height=0.3\textwidth]{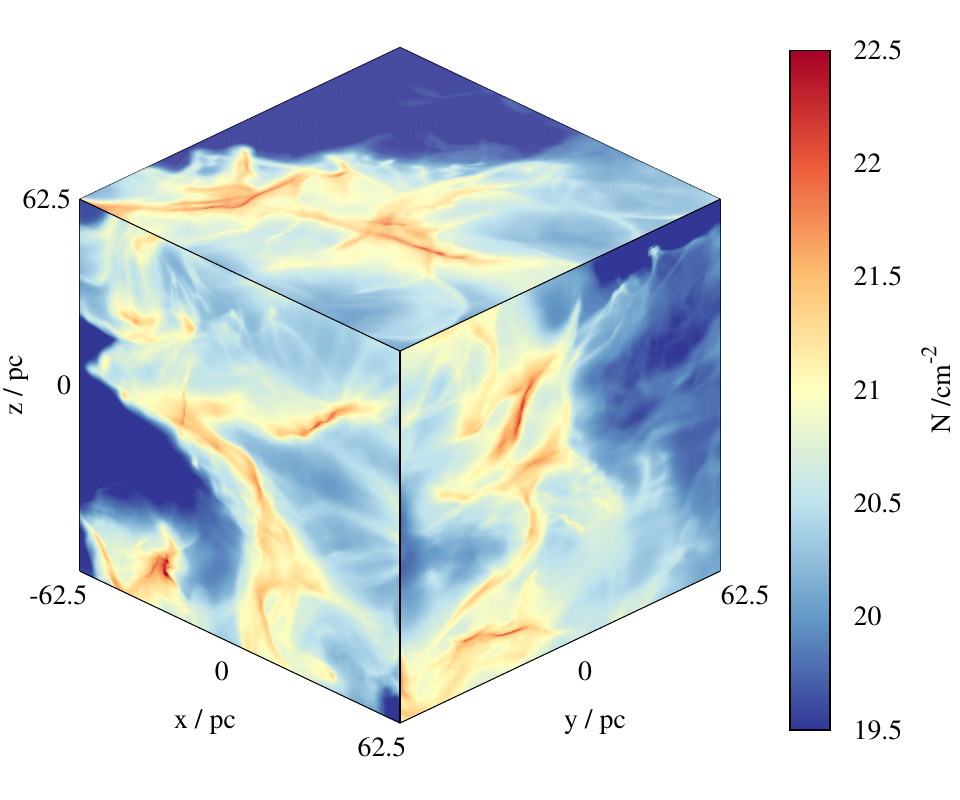}\\
 \includegraphics[height=0.3\textwidth]{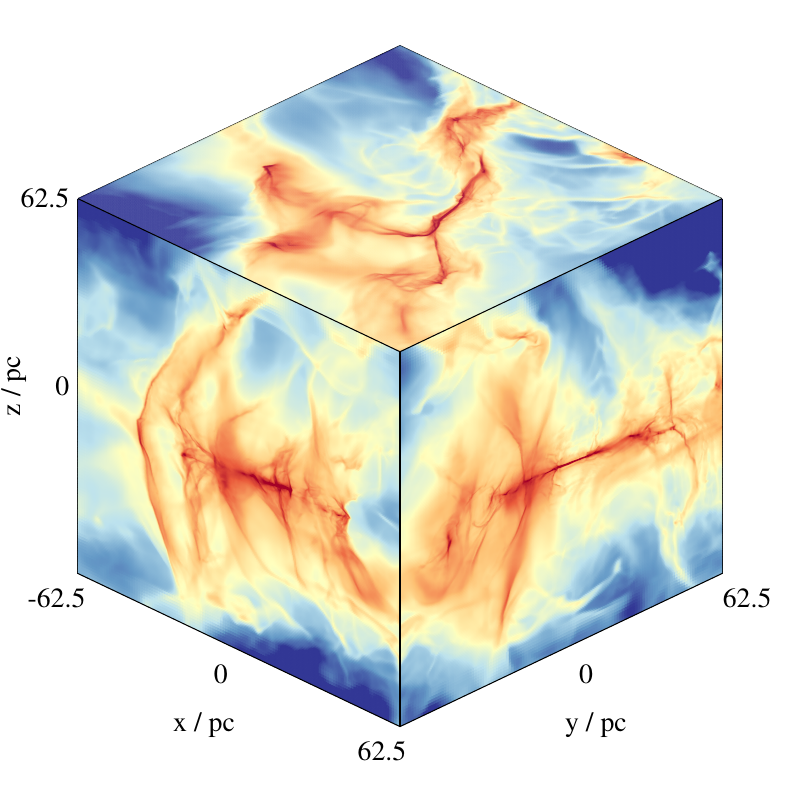}
 \includegraphics[height=0.3\textwidth]{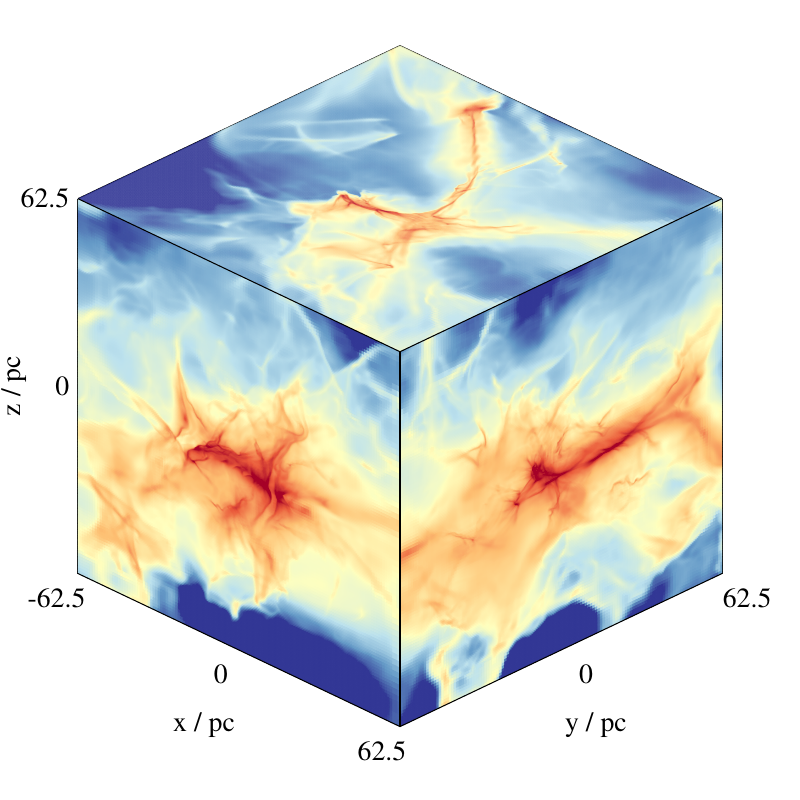}
 \includegraphics[height=0.3\textwidth]{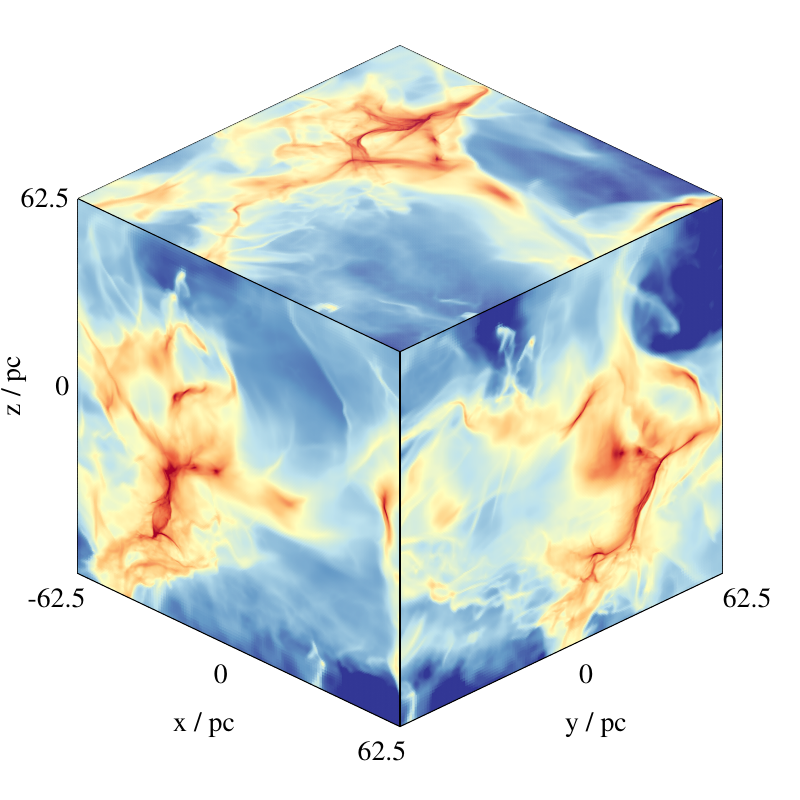}\\
 \caption{Overview of the six different SILCC-Zoom simulations SILCC-MC1 to SILCC-MC6 (from top left to bottom right) at $t_\rmn{evol}$ = 3 Myr showing the column density projected from all three sides around the center of the corresponding zoom-in region.}
 \label{fig:CD_Zoom}
\end{figure*}

Similar to the CF simulations, the times given in the following refer to the time elapsed since the start of the zoom-in procedure at $t_0$ (see Table~\ref{tab:overview}), $t_\rmn{evol}$ = $t$ - $t_0$. As stated in Section~\ref{sec:initial-zoom}, the initial magnetic field strength of the SILCC-Zoom simulations is \mbox{$B_{x,0}$ = 3 $\mu$G}, which is slightly below the threshold value of $\sim$ 5 $\mu$G for which a change in the relative orientation occurred in the CF simulations (see the Fig.'s~\ref{fig:PRS_CF} and~\ref{fig:zeta_CF}). As for the CF simulation we first consider the results in 2D.

In Fig.~\ref{fig:CD_Zoom} we show the column density for the 6 SILCC-Zoom simulations. The simulations show only small changes of the PRS over time, which is why we here consider only the situation at \mbox{$t_\rmn{evol}$ = 3 Myr}. However, as the results strongly dependent on the chosen LOS, we show the column density projected from all three sides. The clouds show a pronounced filamentary structure which is partly shaped by the SNe going off prior to $t_0$. The mass of the gas with $n$ $\geq$ 100 cm$^{-3}$ of the clouds at this time ranges from about $6 \times 10^3$ M$_{\sun}$ (SILCC-MC3) to $56 \times 10^3$ M$_{\sun}$ (SILCC-MC4) \citep[see Fig.~1 in][]{Seifried19}, thus covering the typical range for Galactic MCs \citep[e.g.][]{Larson81,Solomon87,Elmegreen96,Heyer01,Roman10,Miville17}.

In the right panel of Fig.~\ref{fig:pol_maps} we show the polarization degree and direction of run SILCC-MC1. The polarization degree reaches values up to a few 10\% and the polarization pattern indicates a moderately complex magnetic field structure as expected for turbulent environments \citep[see also][for more details on the accuracy of the observed polarization degree and structure]{Seifried19}. The other runs (not shown here) show qualitatively similar results in particular with respect to the pattern of the polarization degree and the structure of the inferred magnetic field.

\begin{table}
\caption{The strength of the magnetic field, $|\left\langle \bf{B} \right\rangle|$, $\left\langle \bf{B} ^2\right\rangle^{1/2}$ and $ \left\langle \bf{B}_\rmn{rand} \right\rangle$ for the SILCC-Zoom simulations at \mbox{$t_\rmn{evol}$ = 3 Myr}.}
\centering
\begin{tabular}{lcccc}
  \hline
 run &  $|\left\langle \bf{B} \right\rangle|$  ($\mu$G)   & $ \left\langle \bf{B}^2 \right\rangle^{1/2}$ ($\mu$G) & $ \left\langle \bf{B}_\rmn{rand} \right\rangle$ ($\mu$G)   \\
 \hline 
 SILCC-MC1 & 2.6 & 4.4 & 3.6   \\
 SILCC-MC2 & 2.1 & 3.6 &  2.9 \\
 SILCC-MC3 & 1.0 & 2.1 & 1.9 \\
 SILCC-MC4 & 2.9 & 4.7 &  3.7 \\
 SILCC-MC5 & 2.7 & 4.6 & 3.7  \\
 SILCC-MC6 & 2.6 & 3.7 & 2.7  \\
 \hline
 \end{tabular}
 \label{tab:bsilcc}
\end{table}

Despite some variations in the magnetic field structure, the two projections perpendicular to the $x$-axis in the right panel of Fig.~\ref{fig:pol_maps} show that the mean field direction is still along the $x$-axis. In order to investigate this more quantitively, we consider the evolution of the mean magnetic field strength, $|\left\langle \bf{B} \right\rangle|$, the total field strength, $\left\langle \bf{B} ^2\right\rangle^{1/2}$, and its random component, \mbox{$\left\langle \bf{B}_\rmn{rand} \right\rangle$ = $\sqrt{ \left\langle \bf{B} ^2 \right\rangle - |\left\langle \bf{B} \right\rangle|^2}$}. This is done for the cubes shown in Fig.~\ref{fig:CD_Zoom}, i.e. using a side-length of 125 pc. As their centers are exactly in the midplane of the galactic disc, i.e. at \mbox{$z$ = 0 pc} (see Table~\ref{tab:overview}), the initial field strength for all six regions is identical at the start of the simulation. Using Eqs.~\ref{eq:rhosilcc} and~\ref{eq:bsilcc}, one can derive \mbox{$|\left\langle \bf{B} \right\rangle|$ = $\left\langle \bf{B} ^2\right\rangle^{1/2}$ = 2.2~$\mu$G.} Note that this average is smaller than \mbox{$B_{x,0}$ = 3 $\mu$G} due to the exponential decrease along the $z$-direction.

In Table~\ref{tab:bsilcc} we list $|\left\langle \bf{B} \right\rangle|$, $\left\langle \bf{B} ^2\right\rangle^{1/2}$ and $\left\langle \bf{B}_\rmn{rand} \right\rangle$ at \mbox{$t_\rmn{evol}$ =  3 Myr}. The first is always smaller than the second, as $|\left\langle \bf{B} \right\rangle|$ describes the ordered field, whereas $\left\langle \bf{B} ^2\right\rangle^{1/2}$ also takes into account the random component of the magnetic field, like e.g. field reversals. The random component, $\left\langle \bf{B}_\rmn{rand} \right\rangle$, alone is comparable to or partly even slightly larger than the ordered field. Overall, however, their difference is rather moderate. Furthermore, for all runs even at \mbox{$t_\rmn{evol}$ = 3 Myr} the largest component of $\left\langle \bf{B} \right\rangle$ is still along the $x$-direction, i.e. the direction of the initial magnetic field (Eq.~\ref{eq:bsilcc}). These results thus agree with our findings that also at later times there exists a preferred direction of $\bf{B}$ along the $x$-axis (Fig.~\ref{fig:pol_maps}).

Furthermore, for most of the runs $|\left\langle \bf{B} \right\rangle|$ has increased slightly over time due to gravitational accretion of the ambient gas -- and thus also magnetic flux -- onto the forming clouds. Only for the run SILCC-MC3 there is a clear decrease present, which can be explained by the accompanying mass loss in the considered cube (compare top right panel in Fig.~\ref{fig:CD_Zoom}). The value of $\left\langle \bf{B} ^2\right\rangle^{1/2}$ has increased in all runs (except run SILCC-MC3) due to the random magnetic field component generated by the passing SN shocks and the ongoing gravitational collapse.

\subsection{2D: The relative orientation between the observed magnetic field and $\nabla N$}
\label{sec:prs_zoom}

\begin{figure*}
 \includegraphics[width=\textwidth]{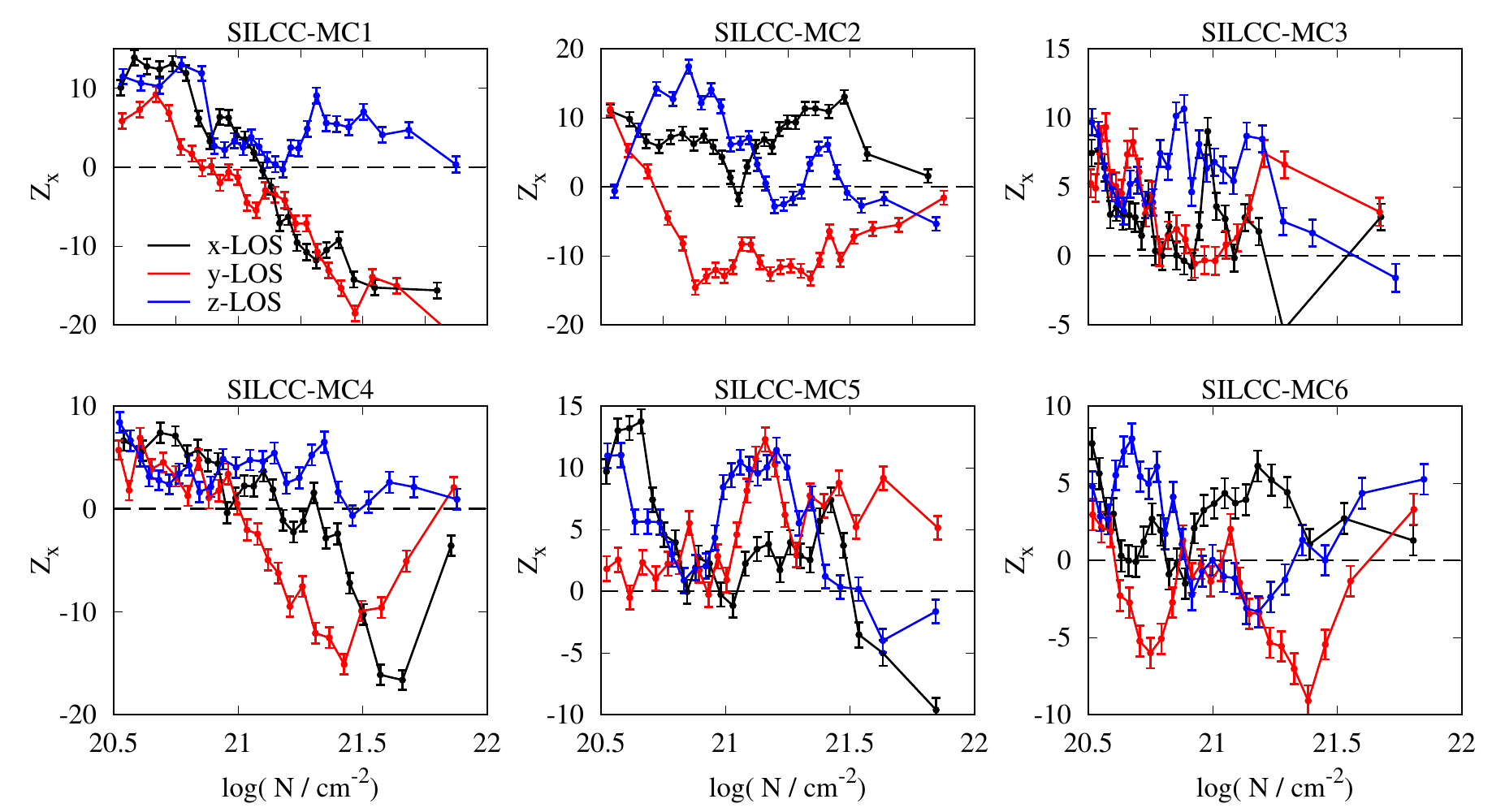}
 \caption{PRS, $Z_x$, of the six SILCC-Zoom simulations at $t_\rmn{evol}$ = 3 Myr for three different LOS. Overall, there is a large variety of shapes in agreement with the moderate field strength of 3 $\mu$G. On average, there appears to be a slight trend of decreasing $Z_x$ with increasing $N$. Note the different ordinate scalings.}
 \label{fig:PRS_Zoom}
\end{figure*}

In Fig.~\ref{fig:PRS_Zoom} we show the PRS, $Z_x$, of the six SILCC-Zoom runs for the three LOS at $t_\rmn{evol}$ = 3~Myr, i.e. for the same snapshots as shown in Fig.~\ref{fig:CD_Zoom}. The polarization and column density maps were smoothed with a Gaussian kernel with a size of 3 pixels before the calculation.

The first thing to notice is the large variety in the shapes of the PRS. The curves show significant qualitative differences between \textit{both} different clouds and between different LOS for individual clouds. Overall, it appears that there is a weak trend of decreasing $Z_x$ with increasing $N$. However, in a number of cases, $Z_x$ does not reach negative values or is -- within the given uncertainty $\sigma_{Z_x}$ -- in agreement with a random orientation, i.e. $Z_x = 0$. In addition, some of the curves show $Z_x \leq 0$ in a narrow range of column densities, before $Z_x$ increases again and then drops towards $\leq 0$ at the highest column densities (e.g. the $x$-direction of SILCC-MC2 and the $y$-direction of SILCC-MC3 and SILCC-MC6).

For some cases (e.g. the $y$-direction of SILCC-MC2, the $x$- and $y$-direction of SILCC-MC4 and the $y$-direction of SILCC-MC6) the PRS reaches negative values at \mbox{$N_\rmn{trans}$ $\simeq$ 10$^{21 - 21.5}$ cm$^{-2}$} and decreases towards higher $N$, but finally increases again towards zero for the highest column densities. Interestingly, this value of $N_\rmn{trans}$ is comparable to that of the CF simulations (see Fig.~\ref{fig:PRS_CF}) and that of actual observations \citep{PlanckXXXV,Jow18,Soler17b,Soler19}. Moreover, it also agrees with that for the transition from sub- to supercritical magnetic fields in the ISM \citep{Crutcher12}.

The trend of random orientation towards the highest column densities is also seen in parts for the CF runs (Fig.~\ref{fig:PRS_CF}). There are three possible explanations for this change from a preferentially perpendicular to a random orientation at very high $N$: first, on the grid scale, the magnetic field structure is not resolved accurately any more due to numerical dissipation/reconnection, thus slightly decoupling from the density and possibly leading to a rather random configuration. This is supported by the fact that the final increase of $Z_x$ appears to happen at lower $N$ for the lower-resolved SILCC-Zoom simulations (0.12 pc) than for the higher resolved CF simulations (0.008 pc). On the other hand, there are also actual observations, which partly show an increase of $Z_x$ towards the highest column densities \citep{Soler17b,Soler19,Pillai20}, which would indicate that the observed increase of $Z_x$ is not due to a limited resolution. Thirdly, also projection effects occurring on very small scales, i.e. high densities, might contribute to an apparent random orientation (see below).

\subsection{3D: The relative orientation of the magnetic field and $n$}
\label{sec:a23_zoom}

\begin{figure}
\includegraphics[width=\linewidth]{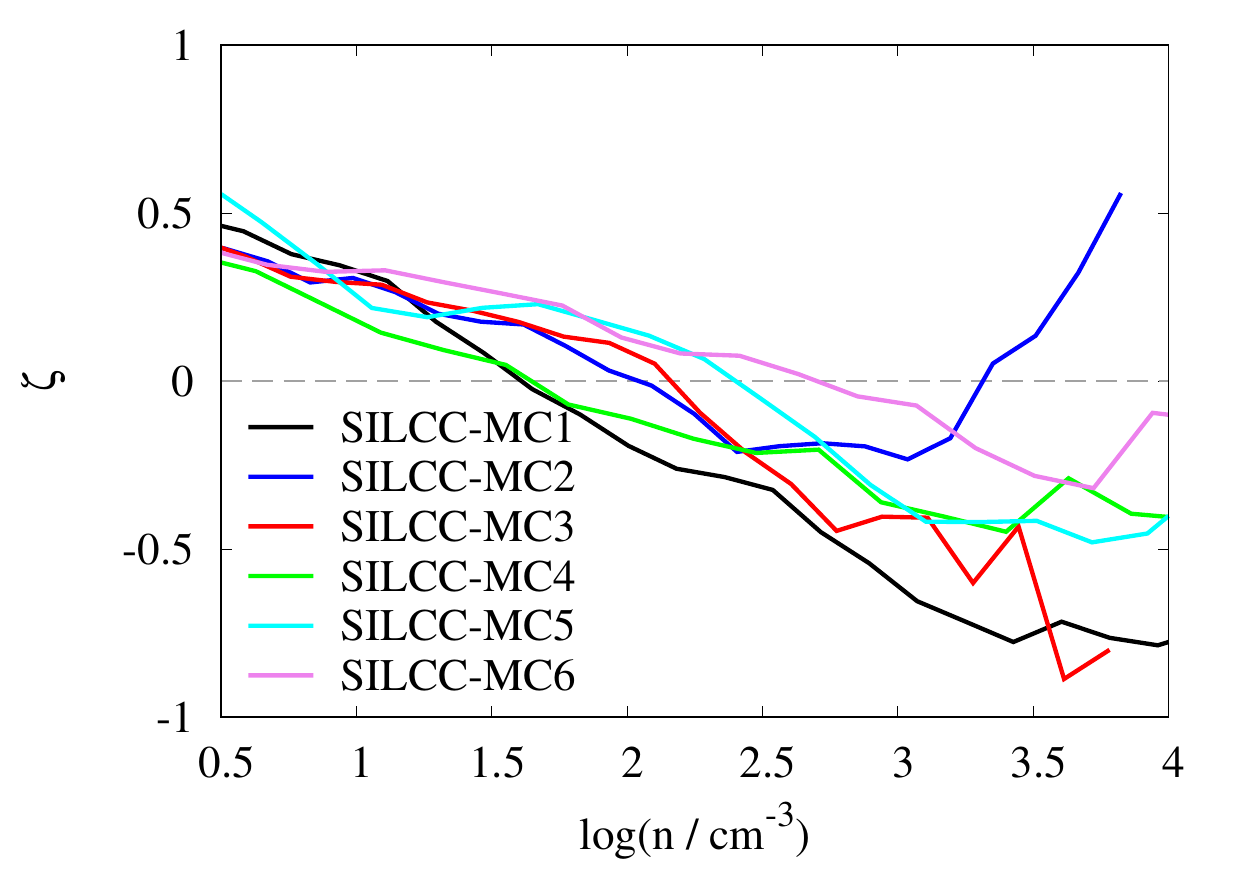}
\caption{Dependence of $\zeta$ on the number density for the SILCC-Zoom simulations at $t_\rmn{evol}$ = 3 Myr. The clear decrease of $\zeta$ with $n$ is in apparent contrast to the 2D analysis (Fig.~\ref{fig:PRS_Zoom}) indicating projection effects which might complicate the analysis in 2D.}
\label{fig:zeta_Zoom}
\end{figure}

\begin{figure}
 \includegraphics[width=\linewidth]{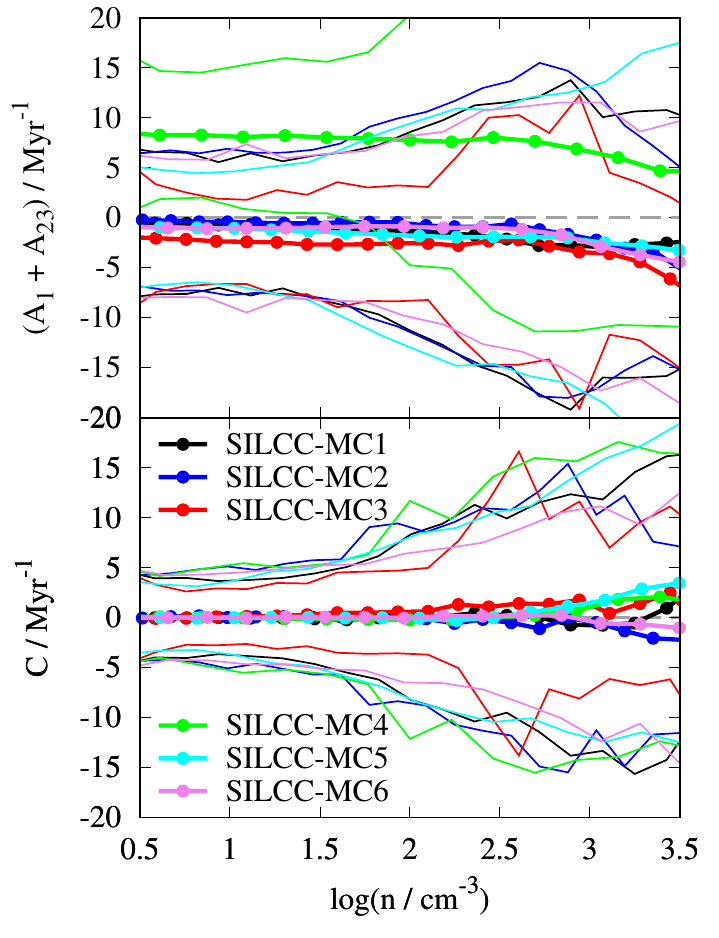} 
 \caption{Density dependence of $A_1$ + A$_{23}$ and $C$ for the six SILCC-Zoom runs at $t_\rmn{evol}$ = 3 Myr. The thick lines with dots show the mean value for a given density, the thin lines the 1$\sigma$ interval. Overall, the width of the distribution is somewhat less than for the CF runs and the values of $A_1$ + $A_{23}$ do not show a clear trend.}
 \label{fig:a23_Zoom}
\end{figure}

The question arises why, for a fixed initial magnetic field strength, the different SILCC-Zoom simulations show such varying PRS. In order to investigate this, we analyse the orientation of the magnetic field and density structures in 3D (Fig.~\ref{fig:zeta_Zoom}). Interestingly, except for run SILCC-MC2, we find a clear trend of decreasing $\zeta$ with increasing density, reaching negative values around \mbox{$n_\rmn{trans}$ $\sim$ 10$^{2 \pm 0.5}$ cm$^{-3}$}. Hence, in 3D the magnetic field shows a perpendicular orientation with respect to the dense structures, which is not clearly visible in the 2D polarization maps (Fig.~\ref{fig:PRS_Zoom}).

This strongly indicates that projection effects can significantly influence -- and thus complicate -- the analysis of the relative orientation in 2D: Observing parallel or random orientations of magnetic fields and column density structures ($Z_x > 0$) does not exclude the possibility that in 3D the magnetic field is oriented perpendicular to the densest structures ($\zeta < 0$). Such projection effects were also reported recently by Girichidis et al. (submitted).

\subsubsection{Comparison with \citet{Soler17a}}

Finally, in Fig.~\ref{fig:a23_Zoom} we also analyse the values of $C$ and $A_1$ + $A_{23}$ in the different zoom-in regions. Except for the run SILCC-MC4, the mean of $A_1$ + $A_{23}$ is always slightly negative ($\sim$ a few -1 Myr$^{-1}$) with a typical spread of $\sim$ 5 Myr$^{-1}$. The individual values of $A_1$ and $A_{23}$ are comparable in size (not shown here). The values of $C$ are on average close to zero except at very high densities and their standard deviation is comparable to that of $A_1$ + $A_{23}$.

On first view the analysis does therefore not present a clear explanation for the observed trend of $\zeta$ (Fig.~\ref{fig:zeta_Zoom}). However, as we show in detail in a semi-analytical analysis of Eq.~\ref{EQ:COSPHI} in Appendix~\ref{sec:appendixA}, a wide distribution of $A_1$ + $A_{23}$ and $C$ around slightly negative mean values can still result in a preferentially perpendicular orientation. Hence, the results of Fig.~\ref{fig:a23_Zoom} can explain the actual orientation of magnetic fields and density structures shown in Fig.~\ref{fig:zeta_Zoom}.  We note that decreasing the width of the distribution or further lowering the mean value increases the probability to find a parallel orientation.

\section{What affects the PRS?}
\label{sec:discussion}

\subsection{Projection effects}
\label{sec:projection}

As discussed in Section~\ref{sec:a23_zoom}, the observed (2D) magnetic field configuration does not necessarily match the actually 3D morphology in the cloud. In fact, the large variety of PRS shapes for the SILCC-Zoom simulations (Fig.~\ref{fig:PRS_Zoom}) agrees well with recent observations of various MCs \citep{PlanckXXXV,Soler17b,Soler19,Jow18,Fissel19}. A compilation of the results of these papers shows that for observed Galactic clouds the full spectrum of PRS shapes found in our simulations is recovered: PRS curves which show (i) local minima, (ii) an increase towards the highest $N$, (iii) and random or perpendicular orientation over the entire column density range. In the context of the results presented here, the clouds in these observations might still have a preferentially perpendicular orientation of the magnetic field and density structures in 3D, though not observed in 2D due to projection effects. This could in particular explain observations of \citet{Soler17b} and \citet{Soler19}, which report significant variations in the PRS when considering different sub-regions of the same molecular cloud complex.

Finally, we note that radiative transfer effects and imperfect dust alignment are unlikely to contribute significantly to these projection effects: As we have shown in \citet{Seifried19}, the observed magnetic field traces well the mass-weighted, LOS-integrated magnetic field. To further support this, we also calculated the PRS using this mass-weighted, LOS-integrated magnetic field instead of that inferred from the polarization maps. We find that the these PRS do not differ significantly from those shown in Fig.\ref{fig:PRS_Zoom}.

\subsection{A critical magnetic field strength}
\label{sec:Bfield}

Overall, our findings of a change from a parallel to a perpendicular orientation of magnetic fields with respect to dense structures with increasing field strength is in good agreement with previous theoretical works \citep[e.g.][]{Heitsch01,Ostriker01,Li04,Nakamura08,Collins11,Hennebelle13,Soler13,Chen15,Chen20,Li15,Chen16,Zamora17,Mocz18}. As pointed out by \citet{Soler17a}, this change of relative orientation is an indicator of compressive motions, i.e. $\nabla \bf{v} <$ 0, coupled with a dynamically important magnetic field. The compressive motions could be created either by converging flows or gravitational collapse. As indicated in the Fig.'s~\ref{fig:CD_CF} and~\ref{fig:PRS_CF}, a stronger magnetic field results in (i) more guided motions towards the central collision interface, (ii) suppressing turbulent motions perpendicular to it, (iii) a stronger gravitation collapse \citep[see also][]{Heitsch09,Zamora18,Iwasaki19} and (iv) consequently a more pronounced perpendicular relative orientation between magnetic fields and the density structures, in good agreement with other theoretical works \citep[][and Girichidis et al. (submitted)]{Soler13,Soler17a,Chen16}.

For the CF runs we find a critical field strength of $\sim$\,5 $\mu$G above which we observe a flip in field orientation. The SILCC-Zoom simulations have an initial magnetic field strength of 3~$\mu$G, which is close to this critical value.  As demonstrated in Section~\ref{sec:a23_zoom}, the observed trend of $\zeta$ in the SILCC-Zoom runs supports the idea of a critical field strength of 3 -- 5~$\mu$G, above which a perpendicular orientation develops.

Interestingly, this value of the initial magnetic field strength is close to the Galactic field strength of about 6~$\mu$G in the solar neighbourhood \citep[e.g.][]{Troland86,Heiles05,Beck13}. It is therefore not surprising that -- also due to possible projection effects -- recently observed Galactic MCs \citep{PlanckXXXV,Soler17b,Jow18,Fissel19} show a similar variety of PRS shapes as the SILCC-Zoom clouds, which have (almost) comparable field strengths.

\subsection{The mass-to-flux ratio}
\label{sec:mu}

As compressive motions during the (later) evolution of MCs are created by gravitational collapse, it appears intuitive to relate the shape of the PRS to the mass-to-flux ratio, which combines the magnetic field strength and gravity in a single parameter. The mass-to-flux ratio is defined as \citep{Mouschovias76}
\begin{equation}
 \mu = \frac{M}{\Phi} \cdot \left(\frac{M}{\Phi}\right)^{-1}_\rmn{crit} = \frac{M}{B \cdot A} \cdot \left(\frac{0.13}{\sqrt{G}}\right)^{-1} \, ,
\end{equation}
where $G$ is the gravitational constant, $A$ and $M$ the area and mass of the cloud, and $\Phi$ = $B \cdot A$ the magnetic flux through it.

For the CF runs we can estimate $\mu$ analytically from the initial velocity, density and magnetic field strength of the inflowing gas (see Section~\ref{sec:initial-CF}) as
\begin{eqnarray}
 \mu_\rmn{CF} =& \frac{2 \times \left((32\, \rmn{pc})^2 \times 13.6\, \rmn{km\, s}^{-1} \times 1.67 \times 10^{-24} \rmn{g\, cm}^{-1} \right) \times \, t}{(32\, \rmn{pc})^2 B_{x,0}} \cdot \left(\frac{0.13}{\sqrt{G}}\right)^{-1} \nonumber \\
 = & 2.85 \times \left( \frac{t}{10\, \rmn{Myr}} \right) \left( \frac{B_{x,0}}{1\, \mu\rmn{G}} \right)^{-1} \, ,
\end{eqnarray}
where the factor of 2 in the nominator accounts for inflow from two sides. In Table~\ref{tab:overview} we list $\mu$ at \mbox{$t$ = 19 Myr} corresponding to \mbox{$t_\rmn{evol}$ = 3 Myr}. At this point, for run CF-B5, i.e. the run with the critical field strength of \mbox{$B_{x,0}$ = 5 $\mu$G}, a value of \mbox{$\mu$ $\simeq$ 1.1} is reached. This is very close to the critical mass-to-flux ratio of $\mu_\rmn{crit}$ = 1, below which gravitational collapse is hampered perpendicular to the magnetic field, but continues unhindered along the field. As stated before, this is in agreement with our simulation results where, for the strong-field (low-$\mu$) cases, the gas appears to be guided along the initial field ($x$-direction) resulting in an accelerated collapse (Fig.~\ref{fig:CD_CF} and Girichidis et al. (submitted)).

Next, we estimate $\mu$ for the various SILCC-Zoom simulations at \mbox{$t_\rmn{evol}$ = 3 Myr}. Using the mass $M$ in the cube with a side-length of 125 pc around the center of the zoom-in region and the mean magnetic field strength given in Table~\ref{tab:bsilcc}, we can approximate the mass-to-flux ratio as
\begin{equation}
 \mu_\rmn{SILCC} = \left(\frac{M}{|\left\langle \bf{B} \right\rangle| \times 125\,\rmn{pc}^2}\right) \cdot \left(\frac{0.13}{\sqrt{G}}\right)^{-1} \, .
\end{equation}
We again find values of $\mu_\rmn{SILCC}$ close to the critical value of 1 (see Table~\ref{tab:overview}, note that the initial value is 1.8). Hence, also for the SILCC-Zoom simulations we expect a flip from parallel to perpendicular orientation to occur matching the results shown in Section~\ref{sec:zoom}.

Moreover, as stated before, the column density of \mbox{$N_\rmn{trans}$ $\simeq$ 10$^{21 - 21.5}$ cm$^{-2}$}, where the transition from parallel to perpendicular orientation occurs in the PRS (Fig.'s~\ref{fig:PRS_CF} and~\ref{fig:PRS_Zoom}), agrees well with the transition point from sub- to supersonic magnetic fields in the ISM \citep{Crutcher12}, which further supports the idea of a connection between both transitions. 

To summarise, we argue that an observed perpendicular orientation of magnetic fields and column density structures indicates a mass-to-flux ratio of $\mu$ $\lesssim$ 1, i.e. very weak magnetic fields can be excluded. Contrary, a parallel orientation indicates $\mu$ $\gtrsim$ 1, excluding the presence of a very strong magnetic field. However, in particular around $\mu$ $\simeq$ 1, projection effects might cause the \textit{observed} relative orientation to be parallel despite the actual, 3D orientation being perpendicular (Section~\ref{sec:zoom}). This limits the PRS analysis to an \textit{exclusion} of a certain range of field strengths. It does, however, not allow a robust determination of $\mu$ and thus the strength of $B$.

\subsection{The column density distribution}

\begin{figure*}
 \includegraphics[width=\linewidth]{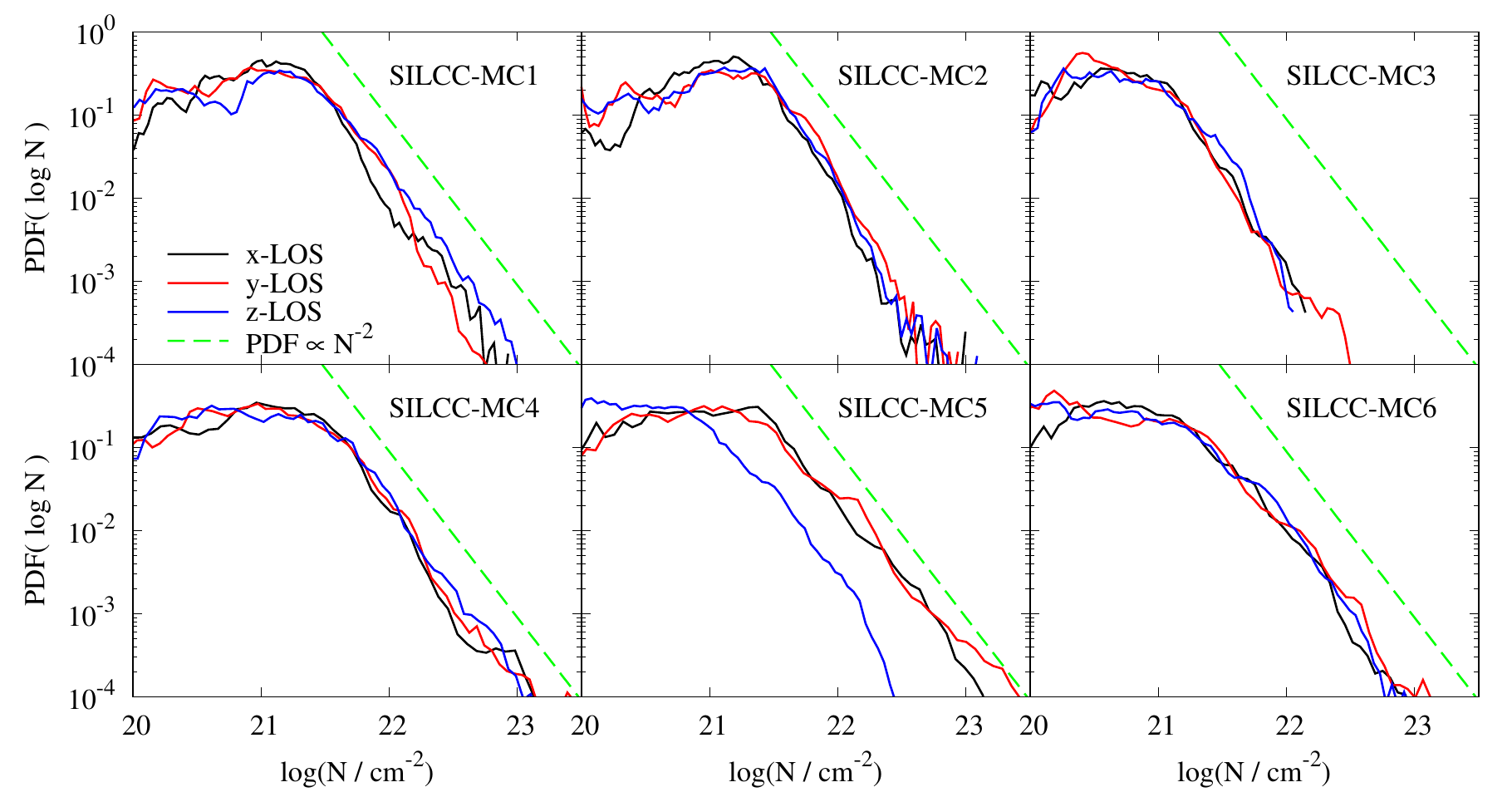}
 \caption{Column density PDF of the different SILCC-Zoom simulations at $t_\rmn{evol}$ = 3 Myr. Overall, the PDFs are relatively similar, which can thus not explain the large variety of shapes of the PRS (compare to Fig.~\ref{fig:PRS_Zoom}). In order to guide the reader's eye, we show a line with a power-law slope of -2 (green dashed line).}
 \label{fig:cd_pdf}
\end{figure*}

Finally, it was suggested by \citet{Soler17b} that the shape of the PRS might be linked to the probability distribution function (PDF) of the column density. In order to check this in the context of our work, we consider the PDFs of log($N$) of the various SILCC-Zoom simulations at $t_\rmn{evol}$ = 3 Myr (Fig.~\ref{fig:cd_pdf}).

The log($N$)-PDFs show the expected, gravity-driven power-law behaviour at high $N$ \citep[e.g.][]{Kainulainen09,Kritsuk11,Girichidis14,Schneider15,Auddy18}. However, for a given run there are only marginal differences in the PDFs at \mbox{$N$ $\gtrsim$ 10$^{20}$ cm$^{-2}$} when considering a different LOS, with the only exception being the run SILCC-MC5. This already indicates that the log($N$)-PDF is not related to the shape of the PRS as e.g. for MC1 the PRS along the $z$-direction differs significantly from those of the other two directions, whereas the log($N$)-PDFs do not show clear variations. Also a comparison of the log($N$)-PDFs of all runs with the corresponding shapes of the PRS does not reveal a coherent picture: The power-law slope of the log($N$)-PDFs for the different runs is around -2~$\pm$~0.5. There is, however, no recognisable correlation of the steepness of the slope with the shape of the PRS, i.e. a higher (or shallower) slope does not result in a particular shape of the PRS (compare to Fig.~\ref{fig:PRS_Zoom}).

\subsection{The impact of resolution and simulation setups}

The CF and SILCC-Zoom runs have significantly different resolutions (0.008 pc vs. 0.12 pc) which might affect their comparison. In order to test a potential impact of different resolutions on the comparability, we repeated run CF-B2.5 with a 4 times (0.032~pc) and 16 times (0.125~pc) lower resolution, i.e. the latter run being comparable to the SILCC-Zoom runs in terms of resolution. We chose run CF-B2.5 as its initial magnetic field properties resemble best that of the SILCC-Zoom runs (see Table~\ref{tab:overview}). As shown in Fig.~\ref{fig:resolution} in Appendix~\ref{sec:appendixB}, the qualitative behaviour of the PRS results is mostly retained despited the difference of a factor of up to 16 in resolution. Moreover, also using differently-sized Gaussian kernels for the calculation of the PRS for the high-resolution CF runs (not shown) does not affect the results significantly. We are therefore confident that the results of the different simulations can be compared to each other. This is also supported by the fact that the SILCC-Zoom simulations and their corresponding magnetic field strengths fit in the trend suggested by the CF runs (Section~\ref{sec:Bfield}).

Furthermore, the CF simulations presented here have an angle of \mbox{$\alpha$ = 0$^\circ$} between the initial magnetic field ($B_{x, 0}$) and the colliding flow along the $x$-direction. It was shown, however, that varying $\alpha$ can significantly affect the formation of dense structures. In particular for high values of $\alpha$, the formation of dense regions is hampered \citep{Heitsch09,Inoue09,Inoue16,Kortgen15,Iwasaki19}. However, \citet{Inoue16} show that when including a realistic level of ISM turbulence, for all values of $\alpha$, the forming low-column density structures ($N$~$\lesssim$~10$^{20.5}$~cm$^{-2}$) are oriented preferentially parallel to the magnetic field\footnote{Note that, as the authors do not include self-gravity, they do not make any statement about structures at higher $N$.}. This is in excellent agreement with the results presented here (Fig.~\ref{fig:PRS_CF}), indicating that our results do not strongly depend on the chosen angle between the initial magnetic field and the colliding flow direction.

This is also supported by the results of the SILCC-Zoom simulations, where the initial SN shocks responsible for forming the clouds do no have any preferred direction with respect to the initial magnetic field. Furthermore, also turbulent box simulations \citep{Heitsch01,Ostriker01,Li04,Collins11,Hennebelle13,Soler13,Li15,Zamora17,Mocz18} show similar results concerning the relative orientation of the magnetic field and gas structures. We thus speculate that the orientation of the initial magnetic field with respect to the (turbulent) flow direction has only a moderate impact on the relative orientation of (column) density structures and the magnetic field. As discussed before, the (observed) orientation is rather influence by the strength of the magnetic field as well as projection effects.

\section{Conclusions}
\label{sec:conclusion}

We present synthetic dust polarization maps of two sets of molecular cloud (MC) formation simulations, colliding flow (CF) simulations and simulations of the SILCC-Zoom project, which models MCs forming from the diffuse, supernova-driven ISM on scales of several 100 pc. The MHD simulations make use of a chemical network and self-consistently calculate the dust temperature by taking into account radiative shielding.

The dust polarization maps are calculated  with the freely available code POLARIS \citep{Reissl16,Reissl19}, which includes a self-consistent treatment of the alignment efficiencies of dust grains with variable sizes. We use radiative torque alignment and present synthetic polarization observations at a wavelength of 1.3~mm. We investigate the simulations concerning the relative orientation of the magnetic field and the density ($n$) structures in 3D and the column density ($N$) structures in 2D. For the latter we apply the Projected Rayleigh Statistics (PRS) introduced by \citet{Jow18}. In the following we summarise our main results:
\begin{itemize}
\item We investigate several CF simulations with increasing magnetic field strength. For these, the analyses of the (observed) relative orientation of the magnetic field in 3D and 2D agree with each other: For magnetic field strengths below $\sim$ 5 $\mu$G, the field has a parallel or random orientation with respect to the $n$- and $N$-structures over the entire range of values.
\item Only for CF runs with strong magnetic fields \mbox{($\gtrsim$ 5 $\mu$G)} a flip from parallel orientation at low values of $n$ and $N$ to perpendicular orientation at high values of $n$ and $N$ occurs. The flip in 3D occurs at $n_\rmn{trans}$ $\simeq$ 10$^3$ cm$^{-3}$ and in 2D at $N_\rmn{trans}$ = 10$^{21 - 21.5}$ cm$^{-2}$. 
\item The SILCC-Zoom simulations all have an initial field strength of 3 $\mu$G and show a flip to a preferentially perpendicular orientation of the magnetic field and filamentary sub-structures at densities \mbox{$n_\rmn{trans}$ $\simeq$ 10$^{2 \pm 0.5}$ cm$^{-3}$}.
\item Based on our results, we suggest that the flip in magnetic field orientation occurs if the cloud's mass-to-flux ratio, $\mu$, is close to or below the critical value of 1. For typical MCs this corresponds to a magnetic field strength around \mbox{3 -- 5 $\mu$G}, which roughly agrees with the strength of the magnetic field in our Galaxy.
\item However, our results clearly demonstrates that projection effects can strongly influence the results of the PRS analysis (in 2D), thus reducing its power to determine the relative orientation of the magnetic field: the observed PRS of the SILCC-Zoom simulations show significant variations among the different runs and different LOS. In case a flip in orientation is present, it typically occurs around $N_\rmn{trans}$ $\simeq$ 10$^{21 - 21.5}$ cm$^{-2}$, but often the column density-based PRS does not show any flip at all.
\item These projection effects can also explain the observed variety in the shape of the PRS, i.e. the magnetic field orientation, of recent observations \citep{PlanckXXXV,Soler17b,Soler19,Jow18,Fissel19}: even if in 3D the relative orientation is preferentially perpendicular, in 2D the postulated flip to a perpendicular orientation at high $N$ might not always be observable. They can also explain the different results obtained for different subregions of an individual MC, e.g. of the Vela C molecular cloud region \citep[][but see also \citealt{Soler19}]{Soler17b}.
\item The column density of $\sim$10$^{21 - 21.5}$~cm$^{-2}$ at which the flip from parallel to perpendicular orientation occurs, agrees well with the transition point from sub- to supercritical magnetic fields in the ISM \citep{Crutcher12}. This further supports the proposed idea of a connection between both transitions.
\item We find that the quantities ($C$, $A_1$ and $A_{23}$), which govern the evolution of the relative orientation in the analytical theory of \citet{Soler17a}, show a wide range of values. We show that their mean values can lead to misleading results in the theory of \citet{Soler17a} and investigate the impact of randomly varying values within the theory. We demonstrate that due to these variations, even slightly negative mean values of $C$, $A_1$ and $A_{23}$ can result in a preferentially perpendicular orientation.
\item Finally, we do not find a correlation between the shape of the PRS and the column density PDF.
\end{itemize}

\section*{Acknowledgements}

The authors like to thank the anonymous referee for the very constructive report which helped to significantly improve the paper. 
DS and SW acknowledge the support of the Bonn-Cologne Graduate School, which is funded through the German Excellence Initiative. DS and SW also acknowledge funding by the Deutsche Forschungsgemeinschaft (DFG) via the Collaborative Research Center SFB 956 ``Conditions and Impact of Star Formation'' (subprojects C5 and C6). SW acknowledges support via the ERC starting grant No. 679852 "RADFEEDBACK".
SR and RSK acknowledge support from  the  Deutsche  Forschungsgemeinschaft via the SFB 881 ``The Milky Way System'' (subprojects B1, B2, and B8) and via the Priority Program SPP 1573 ``Physics of the Interstellar Medium'' (grant numbers KL 1358/18.1, KL 1358/19.2).
RSK acknowledges funding from the Heidelberg Cluster of Excellence {\em STRUCTURES} in the framework of Germany's Excellence Strategy (grant EXC-2181/1 - 390900948).
JDS is funded by the European Research Council under the Horizon 2020 Framework Program via the ERC Consolidator Grant CSF-648 505.
The FLASH code used in this work was partly developed by the Flash Center for Computational Science at the University of Chicago.
The authors acknowledge the Leibniz-Rechenzentrum Garching for providing computing time on SuperMUC via the project ``pr94du'' as well as the Gauss Centre for Supercomputing e.V. (www.gauss-centre.eu).

\section*{Data Availability}

The data underlying this article can be shared for selected scientific purposes after request to the corresponding author.




\bibliographystyle{mnras}
\bibliography{literature} 



\appendix

\section{Equilibrium points of Equation~\ref{EQ:COSPHI}}
\label{sec:appendixA}

As shown in Section~\ref{sec:a23_zoom} for the SILCC-Zoom simulations, the analysis of the mean values of $A_1$ + $A_{23}$ and $C$, which determine the evolution of cos $\varphi$ (Eq.~\ref{EQ:COSPHI}), does not show a clear trend. On first view this disagrees with the finding for $\zeta$ (Fig.~\ref{fig:zeta_Zoom}). However, for all runs the standard deviations of the three parameters are significantly larger than their mean values.

It is of interest if Eq.~\ref{EQ:COSPHI} still possesses some equilibrium points if the parameters $A_1$, $A_{23}$ and $C$ change randomly over time. For this purpose we first aim to rewrite Eq.~\ref{EQ:COSPHI} in a more abstract way. Considering the Fig.'s~\ref{fig:a23_CF} and~\ref{fig:a23_Zoom}, we can see that $A_1$ + $A_{23}$ and $C$ have standard deviations of the order of $\sim$ 10 Myr$^{-1}$. Hence, we can rewrite Eq.~\ref{EQ:COSPHI} to read
\begin{equation}
 \frac{\rmn{d}y}{\rmn{d}t^\ast} = R_1 + R_2 \times y \, ,
\end{equation}
where we use
\begin{eqnarray}
 y & = & \textrm{cos}\, \varphi \nonumber\\
 \textrm{d}t^\ast & = & \textrm{d}t/(0.1\, \textrm{Myr}) \nonumber \\
 R_1 & = & C/(10\, \textrm{Myr}^{-1}) \nonumber \\
 R_2 & = & (A_1 + A_{23})/(10\, \textrm{Myr}^{-1}) \, .
\end{eqnarray}
With this we obtain the linear recurrence relation
\begin{equation}
 y_{n+1} = y_n + (R_1 + R_2 \times y_n) \times \textrm{d}t^\ast = y_n (1 + R_2 \times \textrm{d}t^\ast) + R_1 \times \textrm{d}t^\ast \, .
 \label{eq:y}
\end{equation}
We now determine possible equilibrium points of this relation by evaluating it for many iterations assuming that now $R_1$ and $R_2$ can be described by a stochastic process. For this purpose, we model the time evolution (i.e. many subsequent iterations) of $R_i$ (with $i$ = 1,2) by a stochastic Ornstein-Uhlenbeck process \citep[e.g.][]{Uhlenstein30,Gillespie96}:
\begin{equation}
R_{i,n+1} = R_{i,n} + \frac{1}{\tau_\rmn{OU}}(\mu_\rmn{OU} - R_{i,n}) \times \textrm{d}t^\ast + \sigma_W \textrm{d}W \, .
\label{eq:R}
\end{equation}
Here, $\tau_\rmn{OU}$ is the correlation time of the process, which we choose to be 1 (i.e. 0.1 Myr in physical values\footnote{We note that varying the correlation time does not change the results significantly.}), $\mu_\rmn{OU}$ is the mean (also called drift term), and $\sigma_W$ the standard deviation of the normal distributed Wiener process d$W$. We set
\begin{equation}
\sigma_W = \sigma_\rmn{OU} \times \sqrt{\frac{2}{\tau_\rmn{OU}} \textrm{d}t^\ast} \, ,
\end{equation}
which guarantees that the distribution of $R_i$ (in the limit of an infinitely large number of iterations) has a standard deviation of $\sigma_\rmn{OU}$. For example, in order to obtain a standard deviation of $R_i$ in physical units of 10 Myr$^{-1}$, we set $\sigma_W$ = $\sqrt{2 \textrm{d}t^\ast}$ (as stated before, here $\tau_\rmn{OU}$ = 1).

With this approach, we can now model the movement of a fluid element (note that the time derivative in Eq.~\ref{EQ:COSPHI} denotes a Lagrangian time derivative) through the cloud. As the fluid element moves, $R_1$ and $R_2$ change stochastically (according to Eq.~\ref{eq:R}) on time scales of 0.1 Myr deviating on average by $\sigma_\rmn{OU}$ from a mean of $\mu_\rmn{OU}$.

\begin{figure}
 \includegraphics[width=\linewidth]{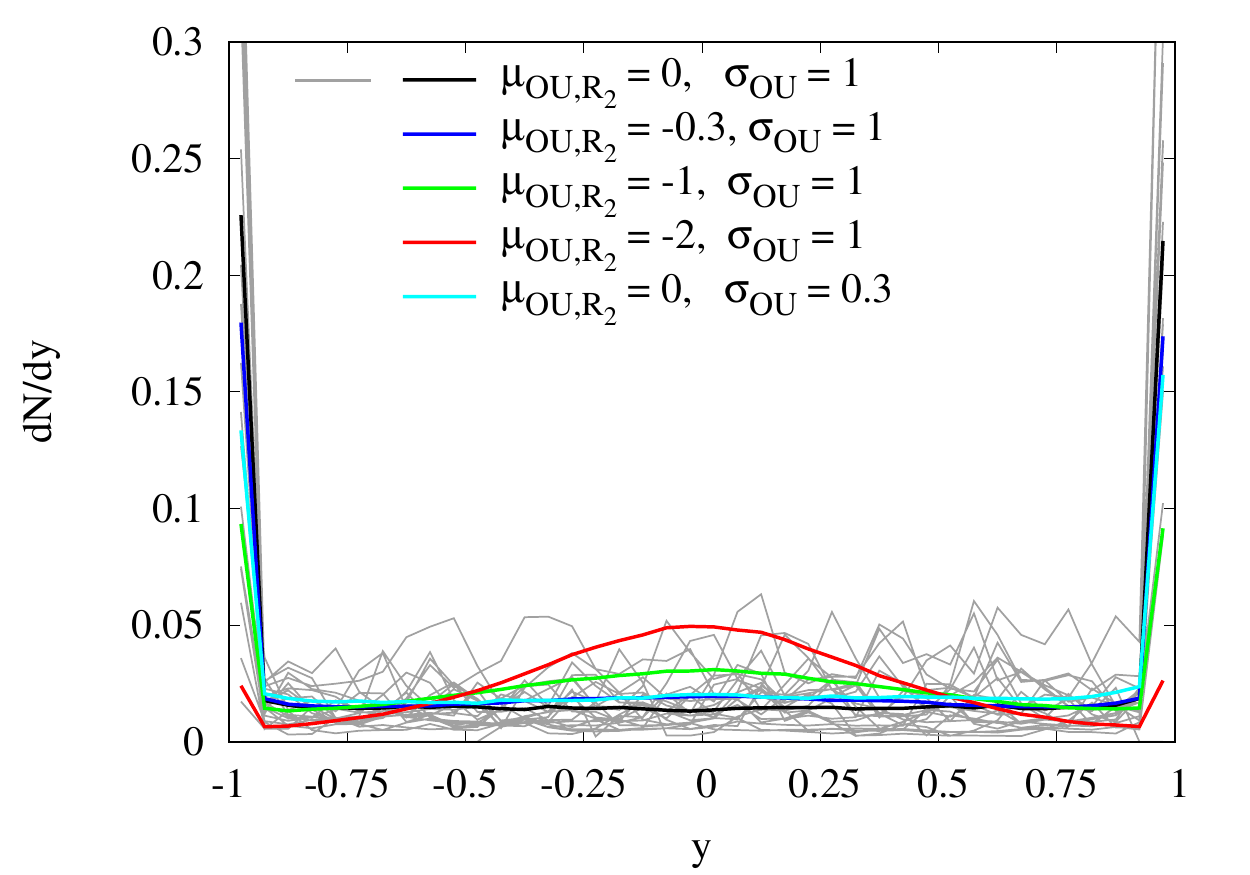}
 \caption{Distribution of $y$ derived for many iterations according to Eq.~\ref{eq:y} where $R_1$ and $R_2$ are modelled by an Ornstein-Uhlenbeck process. Even with $\mu_\rmn{OU, R_2}$ $<$ 0, the distribution shows peaks at $y$ = $\pm$1, which denote the equilibrium points of Eq.~\ref{eq:y}. Physically this can be interpreted as cos $\varphi$ to be close to $\pm$1, i.e. the magnetic field will be preferentially perpendicular to the densest structures. The grey lines show 20 realisations of the case with $\mu_\rmn{OU, R_2}$ = 0.}
 \label{fig:attractor}
\end{figure}

We follow the evolution of fluid elements for 3 Myr, which corresponds to about 30 correlation times. We plot the distribution of $y$ for 20 realisations in Fig.~\ref{fig:attractor} using $\mu_\rmn{OU,R_1}$ = $\mu_\rmn{OU,R_2}$ = 0 and a standard deviation $\sigma_\rmn{OU, R_1} = \sigma_\rmn{OU, R_2} =$ 1 (grey lines). In addition, we show the mean value of the distribution for 1000 of such realisations (black line). As can be seen, there is a strong peak of the distribution at $y$ = cos $\varphi$ = $\pm$1, indicating a preferentially perpendicular orientation of the magnetic field and the density structures (i.e. the magnetic field is parallel to the density gradient).

Next, we change the mean of $R_2$, $\mu_\rmn{OU,R_2}$ to negative values, i.e. the Ornstein-Uhlenbeck will create a sequence of mostly negative $R_2$ values which scatter around $\mu_\rmn{OU,R_2}$. Overall, this should result in $y$ tending towards zero \citep{Soler17a}. However, for both $\mu_\rmn{OU,R_2}$ = -0.3 and -1 (blue and green lines), the distribution shows still pronounced peaks at \mbox{$y$ = $\pm$1}, even though for \mbox{$\mu_\rmn{OU,R_2}$ = -1} there is an increase around $y$ = 0. For $\mu_\rmn{OU,R_2}$ = -2 (red line), however, the distribution is clearly peaked at $y$ = 0. A similar trend can be seen when we decrease the width of the distribution of $R_{1,2}$ from $\sigma_\rmn{OU}$ = 1 to 0.3 and  keep $\mu_\rmn{OU,R_2}$ = 0  (cyan line). Also in this case the peaks at $y$ = $\pm$1 are lower thus resulting in a lower likelihood of finding magnetic fields perpendicular to the density structures.

In order to analyse this more quantitatively, we calculate the probability P($\mathbf{B}$ $\perp$ ISO-$n$) of $\varphi$ to lie between 0$^\circ$ -- 25$^\circ$ and 155$^\circ$ -- 180$^\circ$, i.e. $y$ close to $\pm$1, and the probability P($\mathbf{B}$ $\parallel$ ISO-$n$) of $\varphi$ to lie in the range from 65$^\circ$ -- 115$^\circ$, i.e. $|y|$ $\lesssim$ 0.4. We plot the resulting probabilities in Table~\ref{tab:prob}. For $\mu_\rmn{OU,R_2}$ = 0 and -0.3 it is more likely to find the magnetic field to be perpendicular to the density structures. For $\mu_\rmn{OU,R_2}$ = -1 -- although smaller --  there is still some chance to find the field to be perpendicular to the density structures, which is, however, smaller than P($\mathbf{B}$ $\parallel$ ISO-$n$). In such a case the global cloud average, i.e the PRS, would likely indicated a random or a slightly parallel orientation. Similar holds for the case where the width of the distribution of $R_{1,2}$ is decreased ($\sigma_\rmn{OU}$ = 0.3).

\begin{table}
 \caption{Probability for a preferentially perpendicular orientation of the magnetic field and the density structures (P($\mathbf{B}$ $\perp$ ISO-$n$)) and preferentially parallel orientation \mbox{(P($\mathbf{B}$ $\parallel$ ISO-$n$))} depending on the chosen values of $\mu_\rmn{OU,R_2}$ and $\sigma_\rmn{OU}$ (see text).}
 \begin{tabular}{llcc}
 \hline
   && P($\mathbf{B}$ $\perp$ ISO-$n$)  & P($\mathbf{B}$ $\parallel$ ISO-$n$)  \\
   && ($y$ $\simeq$ $\pm$1) & ($|y|$ $\lesssim$ 0.4) \\
 \hline
 $\mu_\rmn{OU,R_2}$ = 0 & $\sigma_\rmn{OU}$ = 1 & 0.48 & 0.23 \\ 
 $\mu_\rmn{OU,R_2}$ = -0.3 & $\sigma_\rmn{OU}$ = 1 & 0.39 & 0.30 \\
 $\mu_\rmn{OU,R_2}$ = -1 & $\sigma_\rmn{OU}$ = 1 & 0.21 & 0.44\\
 $\mu_\rmn{OU,R_2}$ = -2 & $\sigma_\rmn{OU}$ = 1 & 0.06 & 0.65 \\
 $\mu_\rmn{OU,R_2}$ = 0 & $\sigma_\rmn{OU}$ = 0.3 & 0.33 & 0.30 \\
 \hline
 \end{tabular}
 \label{tab:prob}
\end{table}

This demonstrates that, even if the values of $A_1$ + $A_{23}$ are on average slightly negative, as seen for the majority of the SILCC-Zoom runs (top panel of Fig.~\ref{fig:a23_Zoom}), the wide distribution of $A_1$ + $A_{23}$ and $C$ around their means can cause cos $\varphi$ to be on average close to $\pm$1, i.e. the magnetic field would be preferentially perpendicular to the densest structures. Only for clearly negative mean values or small $\sigma_\rmn{OU}$, as e.g. in the less dense regions (Fig.~\ref{fig:a23_Zoom}), cos $\varphi$ will tend towards 0, i.e. the magnetic field is parallel to the density structures or randomly oriented.

We are aware that the above model is a strong simplification of the actual processes happening in a cloud. However, it allows us to gain some basic insight in the underlying processes governing the relative orientation of magnetic fields and density structures.

\section{Resolution dependence of the CF-B2.5 run}
\label{sec:appendixB}

\begin{figure*}
 \includegraphics[width=0.8\textwidth]{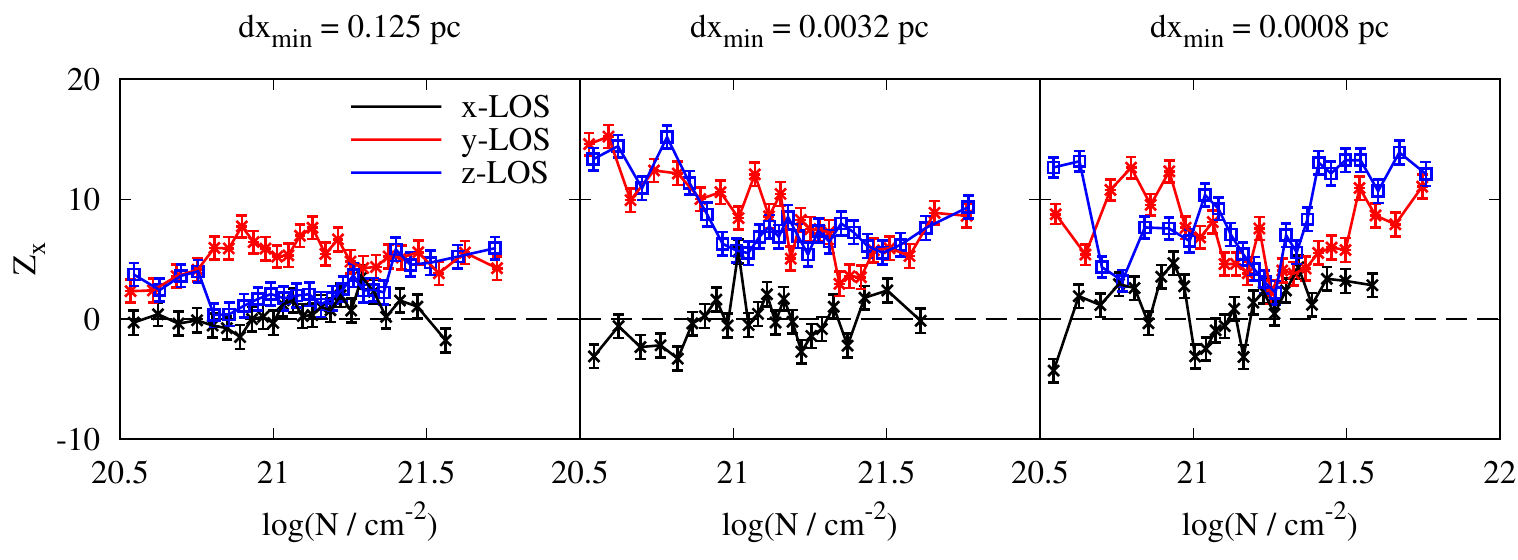}
 \caption{Resolution dependence (from left to right) of the PRS results for the run CF-B2.5. The qualitative behaviour does not change significantly with decreasing resolution, only the statistical significance decreases somewhat.}
 \label{fig:resolution}
\end{figure*}

In Fig.~\ref{fig:resolution} we show the resolution dependence of the PRS results at \mbox{$t$ = 19 Myr} for the run CF-B2.5 simulated with three different maximum resolutions of d$x_\rmn{min}$ = 0.125, 0.0032 and 0.0008~pc. Overall, the qualitative behaviour changes only very little although the resolution changes by a factor of 16. For all resolutions, $Z_x$ remains slightly positive for the $y$- and $z$-direction (indicating a parallel field orientation), whereas for the $x$-direction $Z_x$ $\simeq$ 0. Only the statistical significance decreases with decreasing resolution, i.e. the value of $Z_x$ with respect to the uncertainty $\sigma_{Z_x}$ decreases.

We also compare the mean and total magnetic field strength ($|\left\langle \bf{B} \right\rangle|$ and $\left\langle \bf{B} ^2\right\rangle^{1/2}$, respectively) in the central 32~pc-sized region of the three runs at \mbox{$t$ = 19 Myr}. We find \mbox{$\left\langle \bf{B} ^2\right\rangle^{1/2}$ = 3.3}, 2.9 and 3.0~$\mu$G and \mbox{$|\left\langle \bf{B} \right\rangle|$ = 2.53}, 2.50, and 2.50~$\mu$G for the runs with \mbox{d$x_\rmn{min}$ = 0.0008,} 0.0032 and 0.125~pc, respectively. Hence, even at such a late evolutionary stage, $\left\langle \bf{B} ^2\right\rangle^{1/2}$ differs by $\lesssim$ 10\% and $|\left\langle \bf{B} \right\rangle|$ by even only $\sim$1\%. As the changes in mass are negligible ($\sim$0.1\%), this also implies that the values of $\mu$, which depend on $|\left\langle \bf{B} \right\rangle|$ (see Section~\ref{sec:mu}), show a resolution dependence on a level of a few percent level only.

To summarise, as the initial magnetic field properties of the run CF-B2.5 resemble best that of the SILCC-Zoom runs (see Table~\ref{tab:overview}), the results shown here indicate that the different resolutions of the SILCC-Zoom and CF simulations discussed in the main text do not significantly affect their comparison.

%


\bsp	
\label{lastpage}
\end{document}